\documentclass[a4paper,fleqn]{cas-dc}

\usepackage[numbers]{natbib}

\usepackage{url} 
\usepackage{booktabs}
\usepackage{amsfonts}
\usepackage{nicefrac}
\usepackage{microtype}
\usepackage{amsmath}
\usepackage{lipsum}
\usepackage{mathtools}
\usepackage{caption}
\usepackage{ulem}
\usepackage{subcaption}
\usepackage{subfloat}

\def\tsc#1{\csdef{#1}{\textsc{\lowercase{#1}}\xspace}}
\tsc{WGM}
\tsc{QE}
\tsc{EP}
\tsc{PMS}
\tsc{BEC}
\tsc{DE}

\begin{document}
\let\WriteBookmarks\relax
\def\floatpagepagefraction{1}
\def\textpagefraction{.001}
\shorttitle{DNS of a moderate Re number turbulent evaporating spray}
\shortauthors{WANG et~al.}

\title [mode = title]{Direct numerical simulation of an evaporating  turbulent diluted jet- spray at moderate Reynolds number}                      

\author[1]{Jietuo WANG}[orcid=0000-0002-0422-8024]
\cormark[1]
\ead{jietuo.wang@studenti.unipd.it}

\address[1]{Centro di Ateneo di Studi e Attività Spaziali "Giuseppe Colombo" - CISAS, University of Padova,  Padova (PD) 35131, Italy}

\author[1]{Federico {DALLA BARBA}}%
\ead{federico.dallabarba@phd.unipd.it}

\author[1,2]{Francesco PICANO}%
\ead{francesco.picano@unipd.it}

%

\address[2]{Department of Industrial Engineering, University of Padova, 
Padova (PD) 35131, Italy}

\cortext[cor1]{Corresponding author}

\begin{abstract}
The evaporation of dispersed, liquid droplets in jet-sprays occurs in several industrial applications and in natural phenomena. Despite the relevance of the problem, a satisfactory comprehension of the mechanisms involved has not still be achieved because of the wide range of turbulent scales and the huge number of droplets involved. In this context, we address a Direct Numerical Simulation of a turbulent jet spray at relatively high Reynolds number, $i.e.\ Re=10,000$. We focus on the effect of the jet Reynolds number  on the evaporation process and the preferential segregation of droplets, comparing the outcomes also with a DNS at lower Reynolds number, $Re = 6,000$, in corresponding conditions. The problem is addressed in the hybrid Eulerian-Lagrangian framework employing the point-droplet approximation. Detailed statistical analysis on both the gas and dispersed phases are presented. We found that the droplet vaporization length grows as the bulk Reynolds number is increased from $Re = 6,000$ to $Re = 10,000$ keeping other conditions fixed. We attribute this result to the complex interaction between the inertia of the droplets and the turbulent gaseous phase dynamics. In particular, at higher Reynolds number, the slower droplet mass transfer is not able to comply with the faster turbulent fluctuations of the mixing layer that tend to fasten the process. We also found an intense droplet clustering which is originated by entrainment of dry air in the mixing layer and intensified by the small-scale clustering mechanism in the far-field region. We will show how clustering creates a strongly heterogeneous droplet Lagrangian evolution. All these aspects contribute to the Reynolds-number dependence of the overall droplet evaporation rate.
Finally, we discuss the accuracy of the {\it d-square} law, often used in spray modeling, for present cases. We found that using this law based on environmental conditions the droplet evaporation rate is overestimated.

\end{abstract}


\maketitle

\section{Introduction}
\label{intro}

Droplet-laden, turbulent sprays are complex multiphase flows where a dispersed liquid phase is transported by a turbulent gaseous one. The distinguished phases mutually interact exchanging mass, momentum and energy in the turbulent regime. These complex flows play an important role in many industrial applications as well as in a large variety of natural and environmental processes. Despite the technical and scientific relevance of the problem, the understanding and modeling of the fundamental mechanisms regulating the evaporation and dispersion of droplets in turbulent sprays are still a challenging matter due to the unsteady, multi-scale and multiphase processes involved in these complex flows~\cite{Jenny.2012}. In industry, for instance, an improved understanding of such processes would contribute significantly to the development of high-efficiency and low-emission internal combustion engines. In these applications, the liquid fuel is directly injected into a hot-temperature and high-pressure combustion chamber where the liquid fuel undergoes a phase transition, via primary and secondary atomization, followed by a fierce chemical reaction process within the turbulent gaseous environment. Such a complex process completes in a very short time inside the combustion chamber that multi-scale, multiphase and unsteady phenomena are involved. A typical example is the formation of pollutants which is related to the fluctuations of temperature and reactant concentrations. In particular, soot forms through a pyrolysis process in fuel-rich regions where high temperature subsists without enough oxidizer to react~\citep{Kennedy.1997,Attili.2014,Raman.2016}. This phenomenology can be observed within droplet clusters, where the concentration of the fuel droplets can be even a thousand times higher than its bulk value giving rise to a peak in the fuel vapor concentration.
Another recent instance relies on the transmission of respiratory infections, such as COVID-19, through virus-laden droplets that are formed in the respiratory tract of an infected person and expelled from the mouth and nose during spasmodic events such as coughing and sneezing~\citep{Mittal.2020,Balachandar.2020}. In this context, to prepare ourselves to tackle the outbreak of such respiratory diseases, it is mandatory to improve our knowledge and modeling capability concerning the fundamental mechanisms which govern the dispersion range of virus-laden droplets ejected into the environment within the exhaled turbulent vapor-air mixture.
\par
A phenomenological description of the overall evolution of a turbulent spray can be found in the review of~\citet{Jenny.2012}. The whole process starts with the jet breakup, or atomization, as a high-velocity liquid flow is injected from a duct into a gaseous atmosphere. During the so-called primary atomization, Kelvin-Helmholtz and Rayleigh-Taylor instabilities develop at the gas-liquid interface promoting the jet fragmentation and giving rise to a system of large droplets and ligaments enclosed within the environmental gas~\citep{Marmottant.2004}. Further downstream, secondary atomization follows: the aerodynamic forces, arising as a consequence of the relative inter-phase velocities, causes interface instabilities that lead to a further disintegration of droplets and ligaments into even smaller fragments~\citep{Marmottant.2004,Jenny.2012}. The breakup process occurs in a so-called dense regime and terminates when the surface tension prevails on aerodynamic stresses preventing a further fragmentation~\citep{Jenny.2012}. At this stage, a dilute regime establishes where the dispersed phase volume fraction is small enough that the droplet mutual interactions, such as collisions and coalescence, can be neglected~\citep{Ferrante.2003}. Whereas in the dense regime the vaporization rate is negligible, the vaporization process becomes significant in dilute conditions and most of the liquid evaporates at this stage of the turbulent spray evolution~\citep{Faeth.1995}. In the dilute regime,  small droplets evolve preserving a spherical shape due to the dominance of the surface tension on the aerodynamic stresses~\citep{Elghobashi.1994,Ferrante.2003,Jenny.2012}. Even though the presence of the liquid phase still exerts a considerable effect on the flow in terms of mass, momentum, and energy balance, in dilute conditions, the droplet size is usually below, or comparable to, the dissipative scales of the turbulent flow. In these conditions the point-droplet approximation is widely accepted and the hybrid Eulerian-Lagrangian approach is considered suitable and reliable for the numerical description of problem~\citep{Elghobashi.1994}. In this framework, the mass, momentum, and energy exchanges between the Eulerian carrier phase and Lagrangian point-droplets can be accounted for via distributed sink-source terms in the right- hand side of the balance equations governing the Eulerian phase while the droplet kinematic and thermodynamic variables are evolved in the Lagrangian frame~\citep{Elghobashi.1994,Mashayek.1998,Bukhvostova.2014,DallaBarba.2018,Ciottoli.2020,weiss2020evaporating}.
\par
In order to gain an insight into the fundamental physics of turbulent sprays, various theoretical and numerical studies, which complement traditional experimental techniques, have been conducted with a rapid growth of their applications. One of the former attempts to describe the vaporization process of spherical droplets dragged by a gaseous phase flow relies on the well-known {\it d-square} law, that has been formerly proposed by~\citet{Spalding.1950}~and~\citet{Godsave.1953}. Fixing the thermodynamical properties (temperature, vapor concentration, etc.) in a quiescent environment, they found that the droplet surface decreases linearly with time. Essentially based on the {\it d-square} law, ~\citet{law.1976}~and~\citet{law.1977} proposed the rapid-mixing model: the temperature is assumed to be uniform inside the droplet but temporary varying. The model applies well under the assumption that the thermal conductivity of the liquid phase is high enough to permit a rapid stabilization of the droplet internal temperature in relation to local changes of the environmental temperature. Later, \citet{Abramzon.1989} proposed an improved droplet vaporization model applicable to non-uniform and time-dependent environmental conditions, which takes into account forced convection, molecular diffusion, and the Stefan flow contribution to the vapor transport from the droplet surface to the neighboring environment. The author pointed out that the Ranz-Marshall correlations~\citep{Ranz.1952} for the Nusselt and Sherwood numbers, $Nu$ and $Sh$, respectively, adopted in the rapid-mixing model and {\it d-square} law may lead to a physically incorrect super-sensitivity of the transfer rates to the small turbulent velocity fluctuations in the vicinity of the zero Reynolds number.%
Considering the motion of a finite size spherical particle, or droplet, in turbulent flows,~\citet{Maxey.1983} derived a remarkable equation which accounts for the Stokes drag, added-mass effect and gravitational force under low-Reynolds number circumstance. 
All these approaches, among many other reported in archival literature, have lead to the development of well-established and reliable point-droplet equations in the hybrid Eulerian-Lagrangian frame that have proved to be effective in reproducing droplet evaporation dynamics in both the direct numerical simulation and large eddy simulation frames~\citep{Elghobashi.1994,Miller.1998,Bini.2008,Bukhvostova.2014,weiss2020evaporating,Chatelier.2020}.
\par
Along with the previously reported models for droplet dynamics in turbulent gaseous flows, various numerical techniques have been proposed and many are currently in use to tackle the simulation of dilute sprays laden with dispersed droplets. %
The hybrid Eulerian-Lagrangian numerical approaches based on the Reynolds averaged Navier-Stokes are widely used to perform prediction for industrial applications due to their simplicity and low computational requirements. The main limitation of these methods, which is partially removed in the Large Eddy Simulation frame, is incapable to capture the scales of time-dependent and unsteady motion occurring in turbulent sprays~\citep{Pakhomov.2016}. %
In contrast, the Direct Numerical Simulation (DNS), that resolves all the length and time scales present in a turbulent flow, has proved its capability to capture the whole physics of the spray vaporization process providing an accurate insight into the complex phenomena involved. %
\citet{Mashayek.1998} conducted one of the first DNS of evaporating droplets in turbulent flows. The author neglects the effect of the dispersed phase on the carrier one (one-way coupling) under the assumption of incompressible flow. \citet{Bukhvostova.2014} compared the performances of the incompressible and low-Mach number asymptotic formulation of the Navier-Stokes equation in reproducing the dynamics of a turbulent evaporating spray, considering the effects of changes in the total mass density caused by evaporation of droplets and condensation of water vapor. Although the results don’t show significant differences in mild initial conditions, the usage of the low-Mach number formulation is found to be crucial in order to obtain a reliable quantitative prediction of heat and mass transfer. \citet{Reveillon.2007} studied the impact of preferential segregation of droplets on the mixture fraction field with one-way coupling method. They found that the evaporation process evolves in three different stages in time: single droplet mode in the early stage, cluster mode in the intermediate stage and a gaseous mode in the late stage. Recently,~\citet{DallaBarba.2018} investigated the dynamics of droplet vaporization within a three-dimensional turbulent spatial developing jet in dilute, non-reacting conditions, using DNS framework based on a hybrid Eulerian-Lagrangian approach and the point-droplet approximation. The preferential segregation of droplets is found to be driven by two distinct mechanisms: the inertial small-scale clustering in the jet core and the intermittent dynamics of the jet across the turbulent-non-turbulent interface in the mixing layer, where dry air entrainment occurs. 
\par
Although different numerical studies, in conjunction with theoretical exploitation, have been already conducted, it is still premature to draw a conclusion that a satisfying comprehension of turbulent spray dynamics has been achieved. The capabilities of existing models of reproducing the involved phenomena are not thoroughly confirmed. Moreover, most part of the numerical dataset and studies available in archival literature in the frame of direct numerical simulation is limited to relatively low Reynolds numbers due to the high computational requirements of such simulations. To cover these lacks and to improve model capabilities for applications, the availability of data from three-dimensional direct numerical simulation of evaporating jet sprays at high Reynolds number is necessary. In this context, the present paper aims to address the numerical study of the basic mechanisms governing the evaporation and preferential segregation of droplets in a three dimensional turbulent jet-spray at high Reynolds number. The problem is addressed via a three-dimensional  direct numerical simulation. In particular, the focus is on the effect of the Reynolds number of the jet on different Lagrangian and Eulerian observables, such as evaporation length and droplet size spectrum. %
A comparison between the data produced by the present simulation and the previous DNS data by~\citet{DallaBarba.2018} at $Re=6,000$ is provided. In both cases a similar and strongly inhomogeneous distribution of droplets is observed. Nonetheless, as the Reynolds number is increased from $6,000$ to $10,000$, the vaporization length grows and the mean evaporation rate is lower. 
Different mechanisms providing an explanation for these differences are discussed in detail. The authors believe that the data reported in this paper could contribute to improve our knowledge about the dynamics of turbulent evaporating sprays, providing, at the same time, a  benchmarking test case for the future low-order modeling of the phenomena involved.

\section{Numerical Methodology}
\label{sec:setup}

The numerical results reported in this paper have been computed employing a MPI-parallel code that has been previously used in other numerical studies and has undergone an extensive validation and testing campaign~\citep{Rocco.2015,Picano.2011,DallaBarba.2018,Ciottoli.2020}. The numerical algorithm is based on a hybrid Eulerian-Lagrangian approach and the point-droplet approximation. The exchanges of mass, momentum and energy  between the dispersed phase and the carrier one are accounted for, whereas collisions and coalescence within the dispersed phase are neglected in a two-way coupling frame~\citep{Ferrante.2003}. A brief description of the numerical methodology is provided in the following for the self-consistency of the paper, the reader being referred to the references for additional details~\citep{Rocco.2015,Picano.2011,DallaBarba.2018,Ciottoli.2020}. The conservation equations for the gas phase relies in a low Mach number asymptotic formulation of the Navier-Stokes equations in an open environment:
\begin{align}
&\frac{\partial \rho}{\partial t} + \frac{\partial ({\rho} {u}_j)}{\partial {x_j}} ={S}_m,
\label{eqt:(1)}\\
&\frac{\partial (\rho Y_v)}{\partial t} + \frac{\partial ({\rho} {Y}_{v} u_j)}{\partial {x_j}} = \frac{\partial J_j}{\partial {x_j}} + {S}_m,\label{eqt:(2)}\\
&\frac{\partial (\rho u_i)}{\partial t} + \frac{\partial ({\rho} u_i u_j)}{\partial {x_j}} = \frac{\partial \tau_{ij}}{\partial {x_j}} - \frac{\partial p}{\partial {x_j}} + {S}_{p,i},
\label{eqt:(3)}\\
&\frac{\partial u_j}{\partial x_j} = \frac{\gamma-1}{\gamma} \frac{1}{p_0}\left(\frac{\partial q_j}{\partial x_j} + S_e - L^0_vS_m\right),
\label{eqt:(4)}\\
&p_0 = \rho R_w T,
\label{eqt:(5)}
\end{align}
with
\begin{align*}
J_i = \rho {D}\frac{\partial Y_v}{\partial x_i}, \qquad q_i = k\frac{\partial T}{\partial x_i},
\end{align*}
\begin{align*}
\tau_{ij} = \mu \left(\frac{\partial u_i}{\partial x_j} + \frac{\partial u_j}{\partial x_i}\right) - \frac{2}{3}\mu\frac{\partial u_k}{\partial x_k}\delta_{ij}.
\end{align*}
The Eulerian fields $\rho$, $u_i$, $T$ and $p$ are the density, velocity, temperature and hydrodynamic pressure  of the carrier phase, respectively. In the frame of the low-Mach number asymptotic expansion considered in this paper, the thermodynamic pressure, referred to as $p_0$, results to be spatially uniform and constant in time, due to the open environment conditions~\citet{Majda.1985}. The Eulerian field $Y_v = \rho_v / \rho$ is the vapor mass fraction, with $\rho_v$ the partial density of the vapor. The viscous stress tensor is $\tau_{i,j}$, with $\mu$ the dynamic viscosity of the carrier phase. The parameter $\gamma = c_p/c_v$ is the specific heat ratio of the carrier mixture, where $c_p$ and $c_v$ are its specific heat capacities at constant pressure and volume, respectively. The flux vectors $q_i$ and $J_i$ represent the thermal and mass diffusion and are computed according to the Fourier’s law and Fick’s laws, respectively, the parameters $k$ and $D$ being the thermal conductivity of the carrier mixture and the binary diffusivity of the vapor into the gas.
\par
A reference temperature $T_0 = 0\ K$ is fixed and the assumption of calorically perfect chemical species is considered to estimate the sensible enthalpy. Under this hypothesis, $L^0_v$ is the latent heat of vaporization of the liquid phase evaluated at the reference temperature, $T_0$. The carrier vapor-gas mixture is assumed to be governed by the state equation for perfect gases~\eqref{eqt:(5)}, where $R_w = \mathcal{R} / W$ is the specific gas constant of the carrier mixture, $W$ its molar mass and $\mathcal{R}$ the universal gas constant. Consistently with previous studies in this field ~\citep{Mashayek.1998,Bukhvostova.2014}, the effect of the dispersed phase on the carrier one is accounted for by employing three sink-source coupling terms in the right-hand side of the mass, momentum, and energy equations, $S_m$, $S_{p,i}$,and $S_e$, respectively:
\begin{align}
&S_m = - \sum_{k}\frac{dm^k_d}{dt}\delta(x_i - x^k_{d,i}),
\label{eqt:(6)}\\
&S_{p,i} = - \sum_{k}\frac{d}{dt}(m^k_d u^k_{d,i})\delta(x_i - x^k_{d,i}),
\label{eqt:(7)}\\
&S_e = - \sum_{k}\frac{d}{dt}(m^k_d c_l T^k_d)\delta(x_i - x^k_{d,i}).
\label{eqt:(8)}
\end{align}
The Lagrangian variables $x^k_{d,i}$, $u^k_{d,i}$, $m^k_d$ and $T^k_d$ are the position, velocity, mass and temperature of the droplets, respectively, whereas the parameter $c_l$ is the specific heat of the liquid phase. The summations are taken over the whole droplet population located within the domain (index $k$). The delta function, $\delta(x_i-x_{d_i}^k)$, accounts for the fact that the sink-source terms act only at the Eulerian positions occupied, at a given time, by the point-droplets. In the numerical algorithm, the Eulerian terms~\eqref{eqt:(6)}-\eqref{eqt:(8)} are calculated, in correspondence of each node of the computational grid, by volume averaging the mass, momentum, and energy sources arising from all the droplets located within the grid cell pertaining to the considered grid node.
\par
As mentioned above, the point-droplet approximation is adopted to describe the dispersed phase: the droplets are treated as rigid evaporating spheres and the liquid properties (e.g. temperature) are assumed to be uniform inside each droplet. Since the focus of this paper is on the spray dynamics restricted to dilute conditions, mutual interactions among droplets (e.g. collisions, and coalescence) are neglected. For a discussion about the validity of the preceding assumptions the reader is referred to~\citep{DallaBarba.2018}. Under these hypotheses, the dynamics of the droplets is completely described by the following Lagrangian equations:
\begin{align}
\frac{dx_{d,i}}{dt} &= u_{d,i},
\label{eqt:(9)}\\
\frac{du_{d,i}}{dt} &= \frac{1 + 0.15\ Re^{0.687}_d}{\tau_d} (u_i - u_{d,i}),
\label{eqt:(10)}\\
\frac{dm_d}{dt} &= - \frac{1}{3\tau_d}\frac{Sh}{Sc}m_d\ln{(1+B_m)},
\label{eqt:(11)}\\
\frac{dT_d}{dt} &= \frac{1}{3\tau_d}\left[\frac{Nu}{Pr}\frac{c_{p,g}}{c_l}(T-T_d) - \frac{Sh}{Sc}\frac{L_v}{c_l}\ln(1+B_m)\right],\label{eqt:(12)}
\end{align}
where the parameter $\rho_l$ is the density of the liquid phase, $c_{p,g}$ is the heat capacity at constant pressure of the gaseous component of the carrier mixture and $L_v$ the latent heat of vaporization of the liquid phase evaluated at the droplet temperature. The variable $\tau_d$ is the droplet relaxation time, $\tau_d = 2 \rho_l r^2_d/(9\mu)$, with  $Re_d = 2\rho \| \ u_i - u_{d,i}\| r_d/\mu$ is the droplet Reynolds number. The Schiller-Naumann correlation is adopted in equation~\eqref{eqt:(10)} to account for the effect of the finite Reynolds number of the droplets on the drag. In equations~\eqref{eqt:(11)}~and~\eqref{eqt:(12)}, the Schmidt number, $Sc = \mu / (\rho D)$, and Prandtl number, $Pr = \mu / (c_p k)$ are employed to compute the mass diffusivity and thermal conductivity, respectively. Besides, the Nusselt number, $Nu_0$, and Sherwood number, $Sh_0$, are estimated as a function of the droplet Reynolds number through the Fr\"ossling correlation:
\begin{equation} \label{eqt:(13)}
    Nu_0 = 2 + 0.552Re^\frac{1}{2}_dPr^\frac{1}{3},
\end{equation}
\begin{equation} \label{eqt:(14)}
    Sh_0 = 2 + 0.552Re^\frac{1}{2}_d Pr^\frac{1}{3}.
\end{equation}
A correction is then applied to $Nu_0$ and $Sh_0$ to account for the effect of Stefan flow~\citet{Abramzon.1989}:
\begin{equation} \label{eqt:(15)}
    Nu = 2 + \frac{Nu_0 - 2}{\ln(1 + B_t)}\frac{B_t}{(1 + B_t)^{0.7}}, \quad B_t = \frac{c_{p,v}}{L_v}(T - T_d) 
\end{equation}
\begin{equation} \label{eqt:(16)}
    Sh = 2 + \frac{Sh_0 - 2}{\ln(1 + B_m)}\frac{B_m}{(1 + B_m)^{0.7}}, \quad B_m = \frac{Y_{v,s} - Y_v}{1 - Y_{v,s}}
\end{equation}
where $c_{p,v}$ is the specific heat capacity of the pure vapor at constant pressure, $Y_v$ is the vapor mass fraction evaluated at the droplet position, whereas $Y_{v,s}$ is the vapor mass fraction evaluated at the droplet surface. The latter corresponds to the mass fraction of the vapor, in a fully saturated vapor-gas mixture, evaluated at the droplet temperature. In order to estimate $Y_{v,s}$ we employed the Clausius-Clapeyron relation:
\begin{equation} \label{eqt:(17)}
    \chi_{v,s} = \frac{p_{ref}}{p_0}exp\left[\frac{L_v}{R_v}\left(\frac{1}{T_{ref}} - \frac{1}{T_d}\right)\right]
\end{equation}
where $\chi_{v,s}$ is the vapor molar fraction evaluated at the droplet temperature and $p_0$ the thermodynamic pressure.
The parameter $p_{ref}$ is the saturated vapor pressure evaluated at the reference temperature $T_{ref}$, whereas $R_v = \mathcal{R}/W_l$ is the specific constant of the vapor. The saturated vapor mass fraction is, then,
\begin{equation} \label{eqt:(18)}
    Y_{v,s} = \frac{\chi_{v,s}}{\chi_{v,s} + (1 - \chi_{v,s})\frac{W_g}{W_l}}.
\end{equation}
\par
\begin{figure}
	\centering
    \begin{minipage}[b][10cm][t]{1.\linewidth}
    \vspace*{\fill}
    \centering
    \includegraphics[scale=.3]{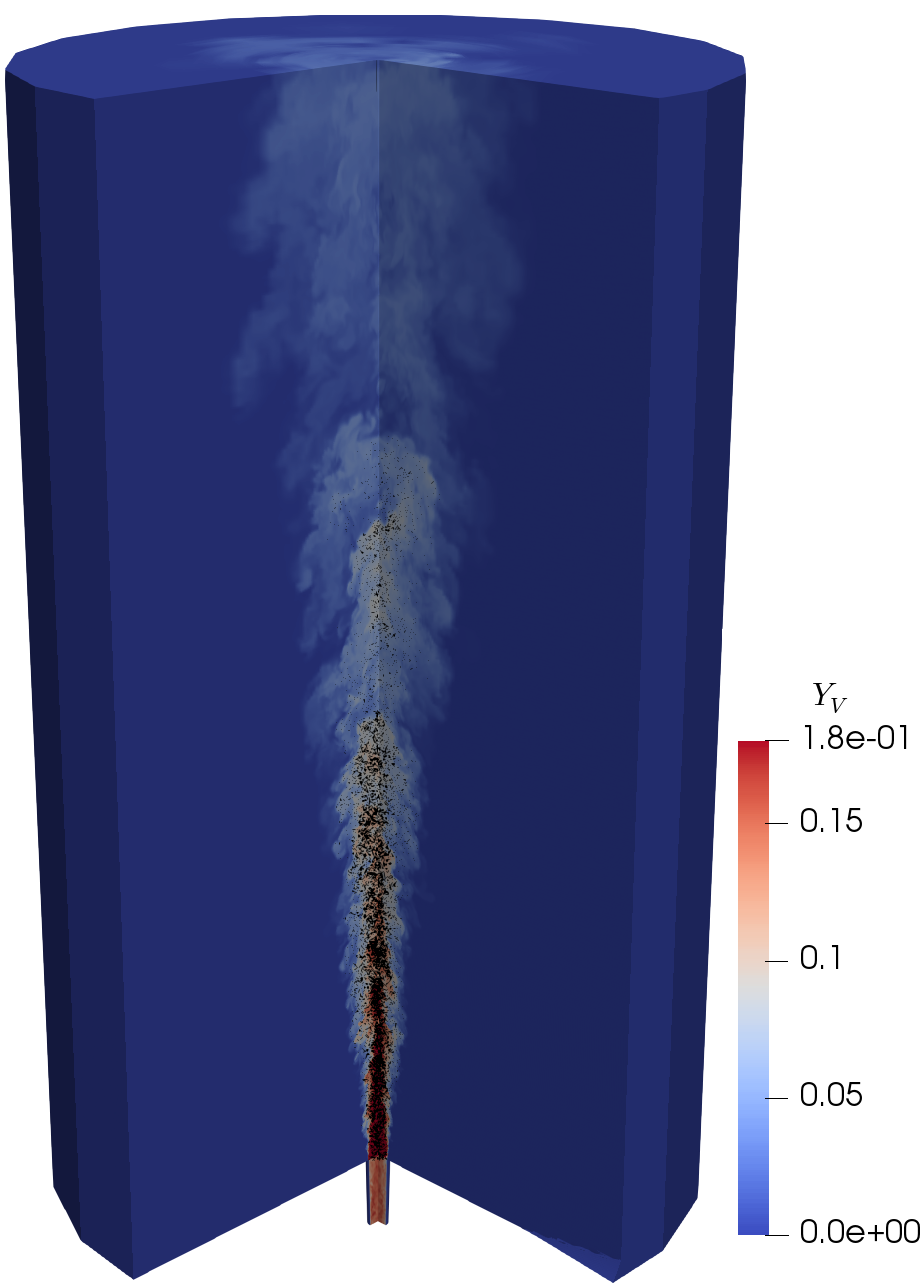}
    \subcaption{3D cylindrical domain}
    \end{minipage}%
    \par\vfill
    \begin{minipage}[b][3.5cm][t]{0.55\linewidth}
    \centering
    \includegraphics[height=3cm]{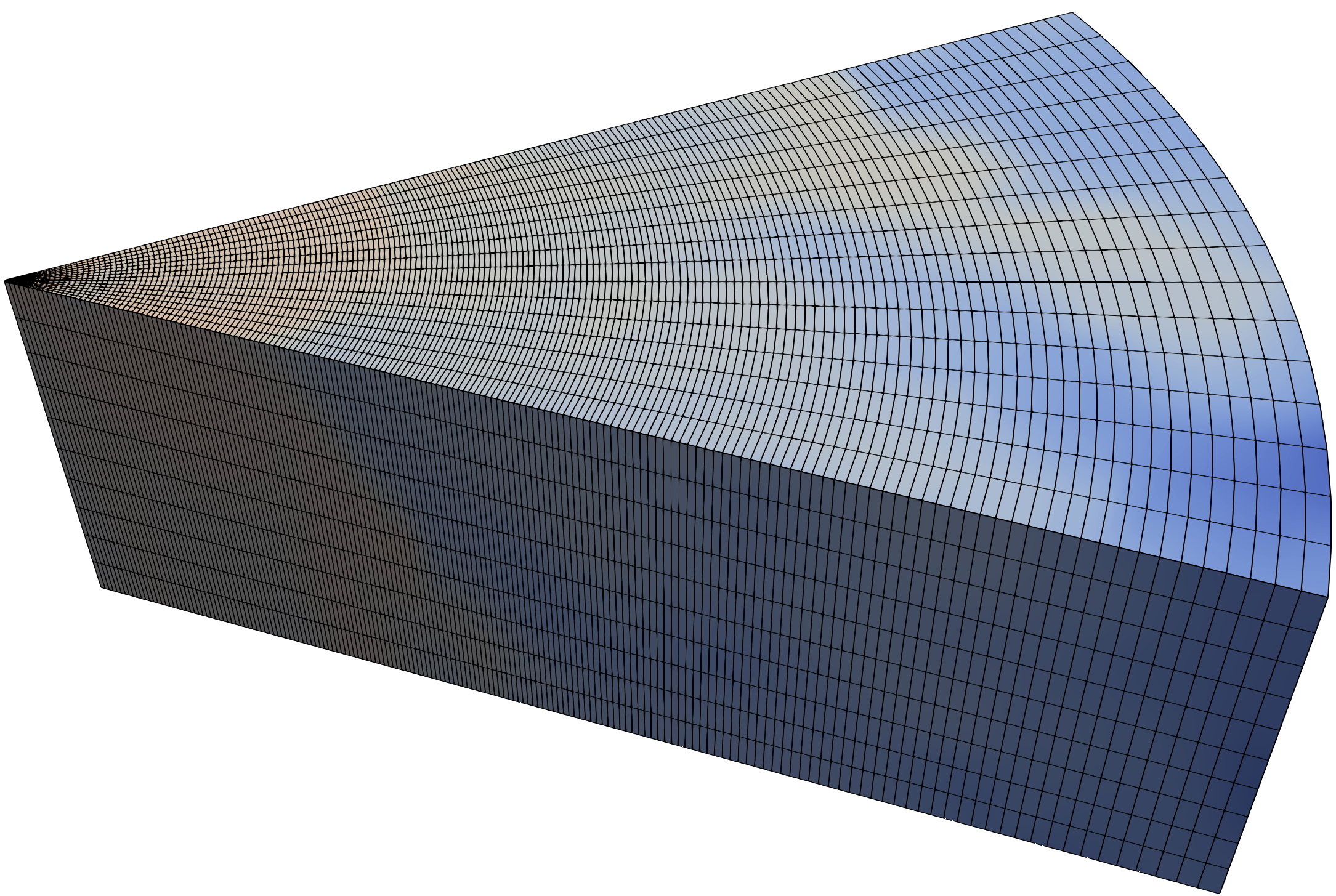}
    \subcaption{Local mesh structure}
    \end{minipage}%
    \begin{minipage}[b][3.5cm][t]{0.4\linewidth}
    \centering
    \includegraphics[height=3cm]{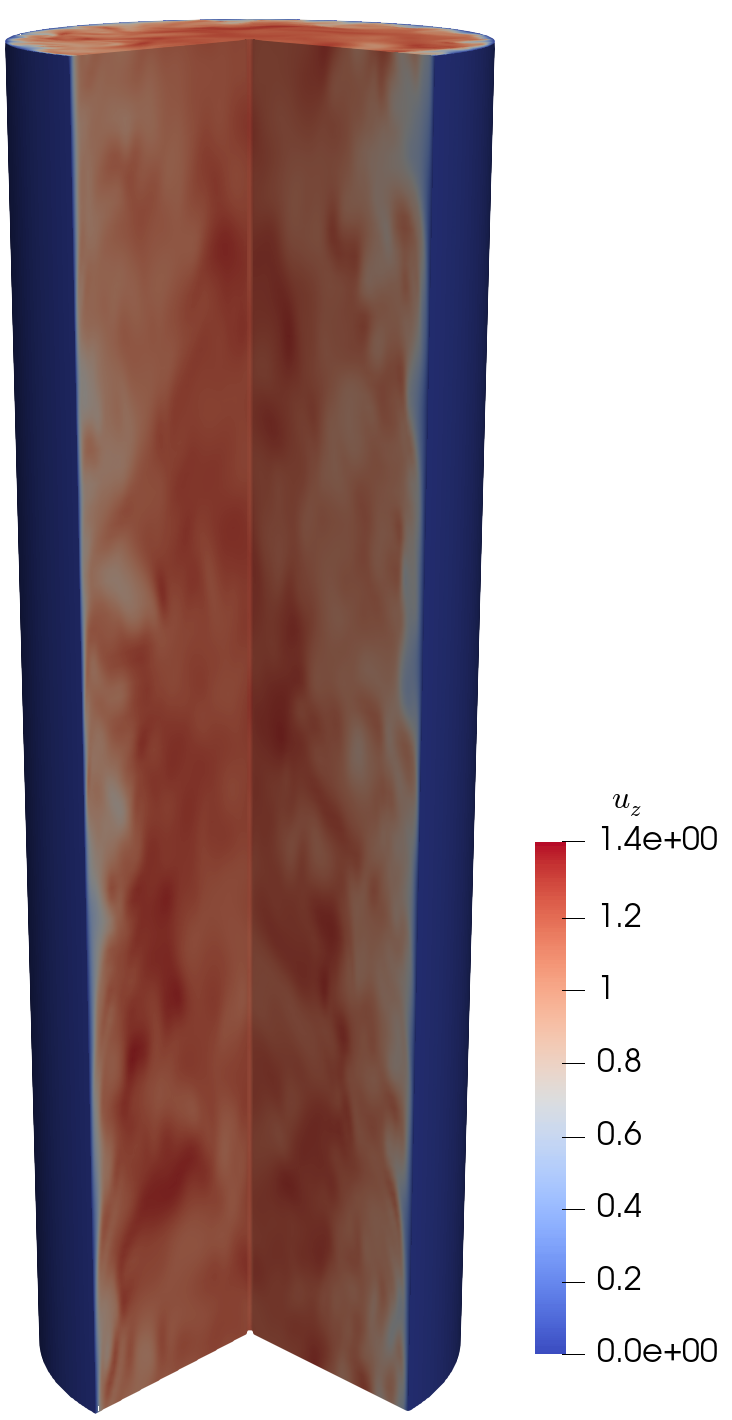}
    \subcaption{3D pipe flow}
    \end{minipage}
	\caption{(a) A snapshot of the computational domain and a representative ensemble of the whole droplet population plotted by black points. (b) A local sector of the mesh structure centered at $z/R=20$. (c) A sketch of the turbulent periodic pipe employed to generate fully turbulent boundary conditions at the jet-spray inflow. The colors contour the vapor mass fraction field, $Y_v$, within the jet and the axial instantaneous velocity, $u_z$, in the turbulent pipe, respectively.}
	\label{fig:dom}
\end{figure}

The Eulerian computational domain is a cylinder. The droplet-laden jet-spray is injected through a circular orifice of radius $R$ located at the centre of the lower base of the domain and streams out towards the opposite one. The domain extends for $2\pi \times 20 R \times 70 R$ in the azimuthal, $\theta$, radial, $r$ and axial, $z$, directions and is discretized by means of a staggered, cylindrical grid of $N_{\theta} \times N_r \times N_z = 192 \times 211 \times 1152$ nodes. An Eulerian algorithm advances in time the Eulerian fields by solving the Low-Mach number formulation of the Navier-Stokes equations~\eqref{eqt:(1)}-\eqref{eqt:(5)}. Second-order, central finite difference schemes are employed for the spatial discretization, whereas the temporal evolution is performed by a low-storage, third-order Runge-Kutta algorithm. The computational grid is uniform along the azimuthal direction whereas it is stretched along the radial and axial ones. The grid spacing is maintained of the same order of the Kolmogorov length scale for the whole downstream evolution of the jet-spray. Details about grid resolution and adequacy of grid spacing can be found in~\citep{DallaBarba.2018} for bulk Reynolds number, $Re=U_0R/\nu=6,000$. In the present case, the grid spacing has been obtained by a scale-up of the computational grid used in~\citep{DallaBarba.2018} in order to maintain an adequate resolution down to the dissipative length scales, but at a higher Reynolds number, $Re=10,000$. The mean grid spacing ratios between the present and reference case, $\overline{\Delta}_{10,000}/\overline{\Delta}_{6,000}$, are about $0.67$ and $0.55$ along the azimuthal and axial directions, respectively. In these directions, the grid spacing ratio is of the same order of the dissipative length-scale ratio $\eta_{10,000}/\eta_{6,000}\simeq(6,000/10,000)^{4/3} \simeq 0.68$. Meanwhile, to get benefits of high-resolution in the shear layer, a stretched and dense mesh instead of a uniform one is implemented along the radial direction near the pipe in which the present mesh spacing is about $0.44$ times of the reference one, though the total grid points don't change too much. A convective boundary condition is adopted at the outlet section located on the upper base of the domain. An adiabatic, traction-free boundary condition is prescribed at the side boundary of the domain making the entrainment of external fluid possible, which, in the present case, consists of dry air. Time-dependent and fully turbulent boundary conditions are prescribed on the inflow section by means of a companion DNS reproducing a fully-developed, periodic pipe flow. A fully turbulent velocity field is assigned on the jet inflow by a Dirichlet condition. This two-dimensional field is computed on a cross-sectional slice of the turbulent pipe. Excluding the circular inflow, the remaining part of the domain base is impermeable and adiabatic. The turbulent pipe extends for $2\pi \times 1 R \times 6 R$ in the azimuthal, $\theta$, radial, $r$ and axial, $z$ directions. The pipe domain is discretized by a staggered grid containing $N_{\theta} \times N_r \times N_z = 192 \times 91 \times 128$ nodes in order to match the corresponding jet-grid nodes at the pipe discharge. A sketch of the cylindrical domain together with the grid structure and the turbulent, periodic pipe is provided in figure~\ref{fig:dom}. Concurrently with the Eulerian solver, a synchronous Lagrangian solver evolves the droplet mass, momentum, and temperature laws~\eqref{eqt:(9)}-\eqref{eqt:(12)} by using the same Runge-Kutta scheme adopted for the Eulerian solver. Numerical stability issues, that can arise when the droplet size becomes too small, are avoided by setting a threshold radius $r_{d,th}$ corresponding to a mass-loss due to evaporation of $99.95\%$ of the initial droplet mass. When the radius of a droplet decreases under the fixed stability threshold, the droplet is considered as completely evaporated and is removed from the simulation.
%
%
\par
The present paper addresses the numerical simulation of liquid acetone droplets dispersed within a gaseous jet consisting of mixture of air and acetone vapor. The jet streams out from the inflow section into an open environment filled by dry air. The absolute pressure of the environment is $p_0 = 101300\ Pa$  while the temperature is set to $T_0 = 275.15\ K$. The radius of the inlet section is set to $R = 4.9 \cdot 10^{-3}\ m$ while the bulk inflow velocity of the jet is $U_0 = 13.9\ m/s$. A monodisperse population of liquid acetone droplets of initial radius $r_{d,0} = 6\ \mu m$ is randomly distributed over the inflow section at each time-step. The injection temperature is fixed to $T_0 = 275.15\ K$ for both the droplets and the carrier mixture. The injection flow rate of the gaseous phase is kept constant by fixing a bulk Reynolds number $Re = 2 U_0 R/\nu = 10,000$, with $\nu = 1.35 \cdot10^{-5}\ m^2/s$ the kinematic viscosity. A nearly saturated condition is prescribed at the inflow for the air-acetone vapor mixture, $S = Y_v/Y_{v,s} = 0.99$, with $Y_v$ the actual vapor mass fraction and $Y_{v,s}(p_0,T_0)$ the vapor mass fraction at saturation, evaluated at the actual inflow temperature and thermodynamic pressure. The acetone mass flow rate is set by means of the ratio $ \Phi = \dot{m}_{act}/ \dot{m}_{air} = 0.28$, where $\dot{m}_{act} = \dot{m}_{act,l} + \dot{m}_{act,v} $ is the sum of the liquid and vapor acetone mass flow rates, while $\dot{m}_{air}$ is the dry-air flow rate. The correspondent bulk volume fraction of the liquid phase is $ \Psi = 8.0 \cdot 10^{-5}$. All the thermodynamic and physical properties of the vapor, gas, and liquid phases are reported in table~\ref{tab:param}. The thermodynamic conditions at the inlet section are comparable to that adopted in the well-controlled experiments on dilute coaxial sprays published by the group of Masri and coworkers~\citet{Chen.2006b}. The time step is set to $\Delta t / t_0 = 0.001 $ where the reference time scale is $t_0 = R / U_0 = 3.5 \cdot 10^{-4} s $. To achieve the prescribed mass flow rate, around $33$ acetone droplets are randomly distributed over the inflow section at each step of the temporal integration. The injection velocity of each droplet is set to be equal to the local velocity of the turbulent carrier phase. The simulation is initialized considering only the single-phase flow until statistical steady conditions have been attained (about $150R/U_0$ time scales). From this step on, droplets are continuously injected and the simulation is run for about $200R/U_0$ time scales in order to achieve a statistical steady condition for the two-phase evaporating flow before collecting data. All the statistics presented in the following are computed considering around $100$ samples separated in time by $R/U_0 = 1$. Concerning the reliability of the simulation, additional information and validation benchmarks about the turbulent periodic pipe, single-phase jet simulations as well as evaporation model can be found in~\citep{DallaBarba.2018,Ciottoli.2020}.

\begin{table}[width=.9\linewidth,cols=4,pos=h]
\caption{Thermodynamic and physical properties of acetone and dry air.}\label{tab:param}
\begin{tabular*}{\tblwidth}{@{} LL|LL@{} }
\toprule
$R$ & $0.0049\ m$& $W_g$ & $0.0290\ kg/mol$\\ 
$p_0$ & $101300\ Pa$ & $W_l$ & $0.0581\ kg/mol$ \\ 
$T_0$ & $275.15\ K$ & $k_g$ & $0.0243\ W/(m \cdot K)$ \\ 
$\mu$ & $1.75 \cdot 10^{-5}\ kg/(m\ s)$ & $k_l$ & $0.183\ W/(m\ K)$ \\
$c_{p,g}$ & $1038\ J/(kg\ K)$ & $D$ & $1.1\cdot 10^{-5}\ m^2/s$ \\ 
$c_{p,v}$ & $1300\ J/(kg K)$ & ${\rho}_l$ & $800\ kg/m^3$\\ 
$c_l$ & $2150\ J/(kg\ K)$ & $L_v$ & $530000\ J/kg$ \\ 
$U_0$ & $13.9\ m/s$ & $r_d$ & $6\cdot 10^{-6}\ m$\\ 
$t_0$ & $3.5\cdot 10^{-4}\ s$ & { } & { } \\ 
\bottomrule
\end{tabular*}
\end{table}

\section{Result and Discussion}
\label{sec:res}
In this section the outcomes from the present DNS case, $Re = 10,000$, is shown and compared with results at $Re=6,000$~\citep{DallaBarba.2018}. The essential difference between the two cases taken into consideration is the bulk Reynolds number of the injected carrier phase, all of the other parameters being the same. From physical point of view, we are comparing two jet sprays with different bulk velocities. It is worth noting that, the initial Stokes number of droplets injected into the domain is larger in the $Re=10,000$ case, since
\begin{equation}
St_0=\frac{2}{9}\frac{\rho_g}{\rho_l}\left(\frac{r_{d,0}}{R_0}\right)^2 Re.
\end{equation}
In particular, the value of the initial Stokes numbers are $St_0\simeq 1.04$ and $St_0\simeq 0.62$ for the $Re=10,000$ and $Re=6,000$ cases, respectively.
\par
A macroscopic overview of the turbulent jets can be given visualizing coherent structures and enstrophy. Several methods have been proposed in order to identify, quantify and visualize the three-dimensional coherent structures in incompressible turbulent flows~\citep{Hunt.1988,Chong.1990,Jeong.1995,Zhou.1999}. We address the visualization of vortical structure by means of the $Q$-criterion~\citep{Hunt.1988}:
%
\begin{figure}
\centering
\begin{subfigure}[b]{1.0\linewidth}
\centering
\includegraphics[width=0.9\linewidth]{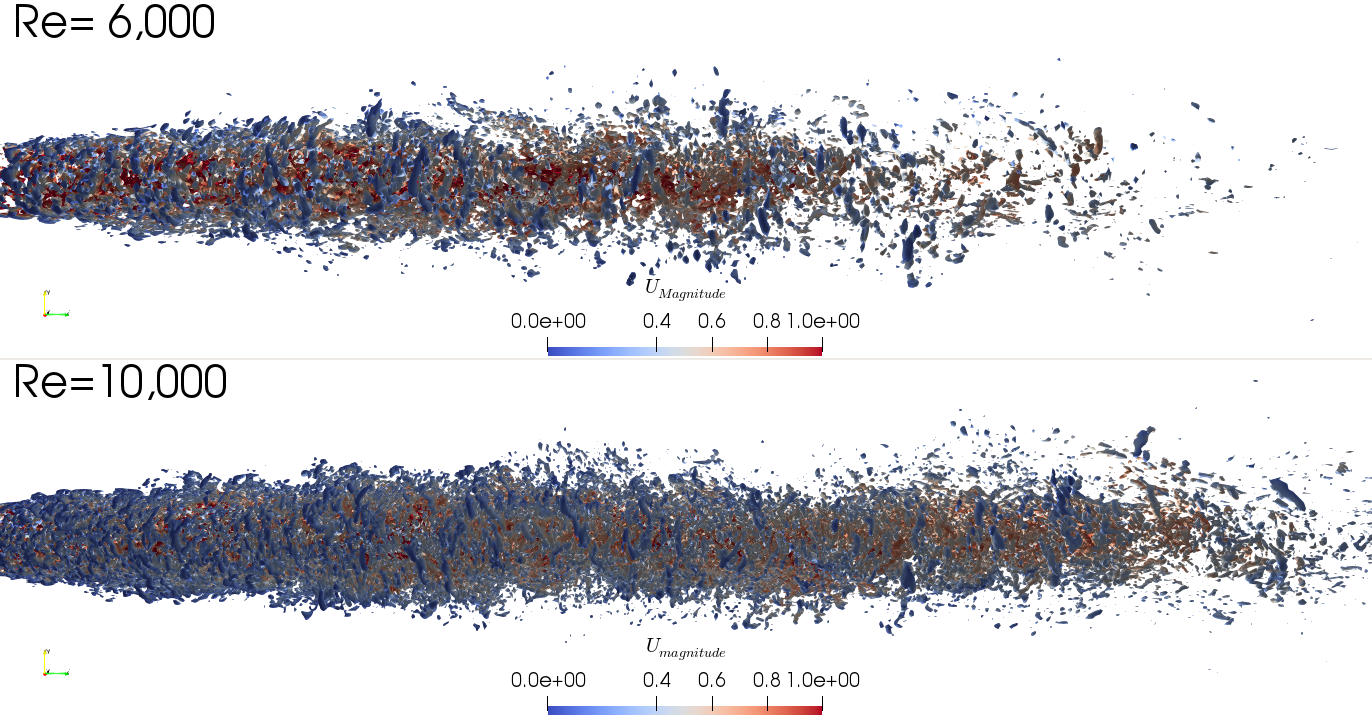}
\caption{}
\label{fig:Vor_a}
\end{subfigure}
\hfill
\begin{subfigure}[b]{1.0\linewidth}
\centering
\includegraphics[width=0.9\linewidth]{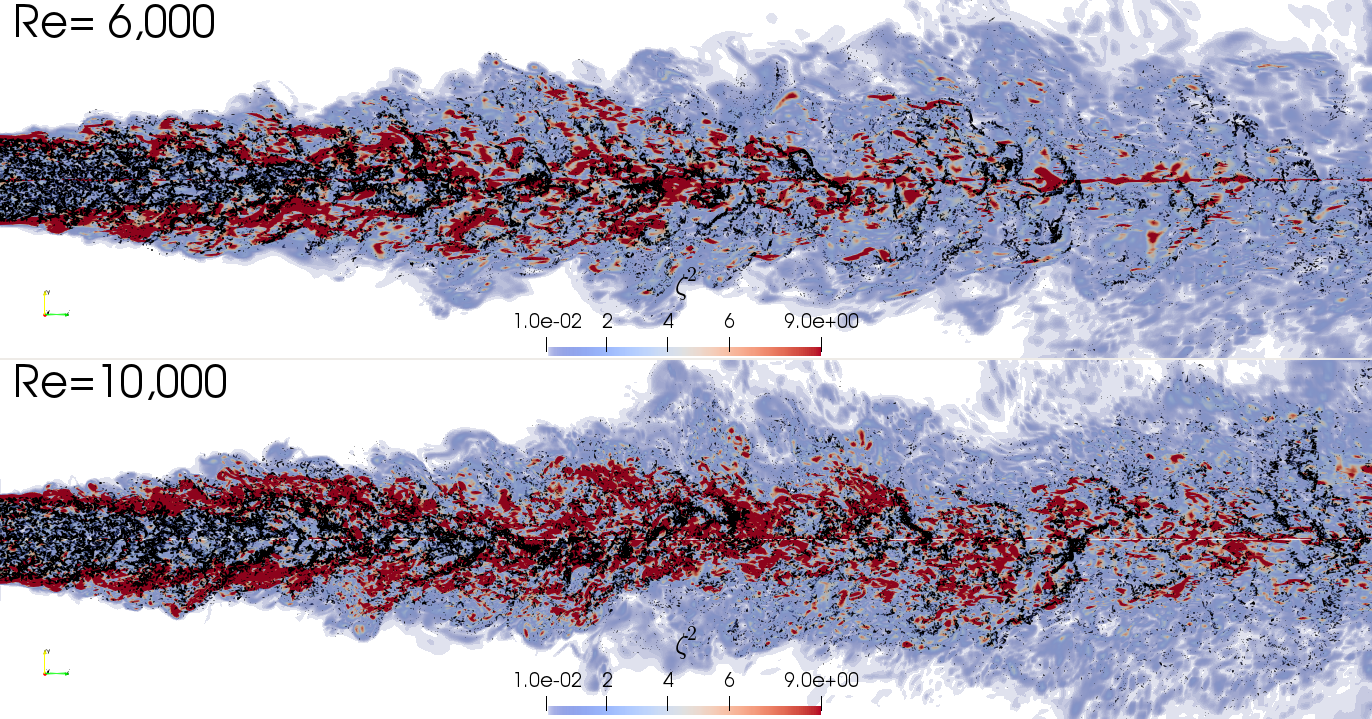}
\caption{}
\label{fig:Vor_b}
\end{subfigure}
\hfill
\begin{subfigure}[b]{1.0\linewidth}
\centering
\includegraphics[width=0.9\linewidth]{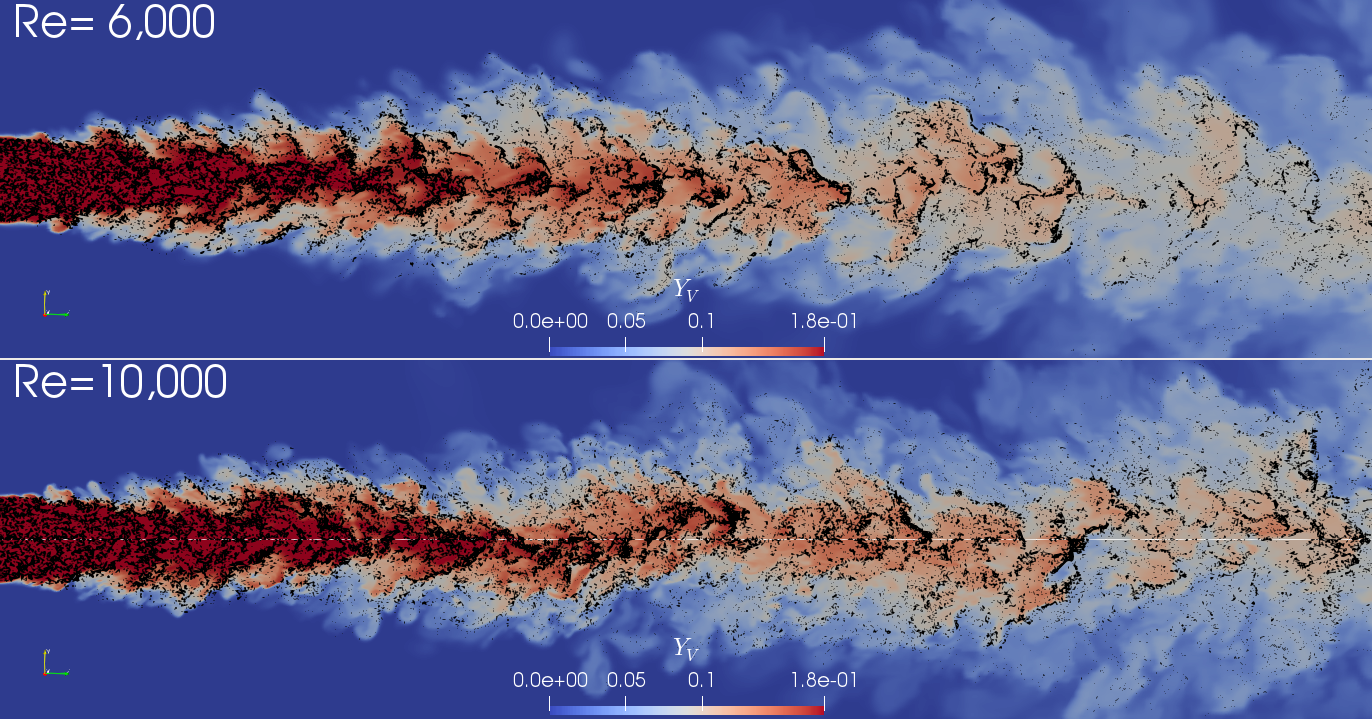}
\caption{}
\label{fig:Vor_c}
\end{subfigure}
\caption{(a) Three-dimensional coherent structures of the flow visualized by the Q-criterion. The iso-surfaces $Q = 1$ are contoured according to the magnitude of the carrier phase velocity. (b) Axial-radial slices of the magnitude of the instantaneous enstrophy field colored using a log scale, $\zeta^2\ =\ \Vert \nabla \times \protect\overrightarrow{u} \Vert$. (c) Radial-axial slices of the turbulent sprays at $Re=10,000$ and $Re=6,000$. The black points represent a subset of the whole droplet population constituted by droplets located within a distance $\Delta/R = 0.1$ to the slice plane. Each point size is proportional to the corresponding droplet radius with a scale factor $100$. The carrier phase is contoured according to the instantaneous vapor mass fraction field, $Y_v$, which is bounded between $0$ and $0.18$, the former value corresponding to the dry condition while the latter to the $99\%$ saturation level prescribed at inflow section. }%
\label{fig:Vor}
\end{figure}
%
\begin{align}
Q = \frac{1}{2}(\Omega_{ij}\Omega_{ij} -  S_{ij} S_{ij}),
\end{align}
with
\begin{align*}
\Omega_{ij} = \frac{1}{2}\left[\frac{\partial u_j}{\partial x_i} - \frac{\partial u_i}{\partial x_j}\right], \quad S_{ij} = \frac{1}{2}\left[\frac{\partial u_j}{\partial x_i} + \frac{\partial u_i}{\partial x_j}\right],
\end{align*}
where $\Omega_{ij}$ and $S_{ij}$ are the vorticity tensor and the rate-of-strain tensor, respectively. By this approach, a vortex is defined as a spatial region where $Q\ >\ 0$, $i.e.$ the Euclidean norm of the vorticity tensor dominates that of the rate of strain~\citep{Haller.2005}. In general, larger $Q$ values are associated with more intensive vortexes. 
Figure~\ref{fig:Vor_a} shows a comparison of the instantaneous vortical structure of the flow for the two Reynolds number cases visualized by the iso-surfaces $Q=1$ contoured by the magnitude of the Eulerian velocity field. Moving downstream from the inflow section, the vortical structures spread out into the far-field while subjected to a decay process. As the Reynolds number is increased, vortical tubes becomes denser and smaller, but also the vortices spread is prolonged downstream~\citep{Ault.2015}. This aspect suggests a longer turbulent spray jet evolution for the higher Reynolds number. 
Figure~\ref{fig:Vor_b} shows a two-dimensional radial-axial snapshot of the instantaneous distribution of droplets and the instantaneous enstrophy field $\zeta^2\ =\ \Vert \nabla \times \overrightarrow{u} \Vert$. As expected, in the near-field, high-intensity vorticity regions are mainly located in the shear layer  whereas vorticity is lower in the jet core. It should be noted that present jets are generated by fully turbulent pipe flows and so fluctuations are present also in the core. Hence the so-called potential core region should be named as unperturbed core region where the turbulence statistics remain closer to the inlet pipe flow conditions and not to those of jet far-field. Moving downstream, the jet decay leads to a different behavior. This dynamics, that is intrinsically related to the physics of turbulent jets, is observed in both the cases, the high-Reynolds jet presenting vorticity-related features that are shifted downstream compared to the lower Reynolds one. The decay of turbulent structure of the jet observed in this paper in consistent with the experimental observation by~\citep{Lau.2014,Lau.2016}. 
Droplets develop clustering and appear less frequently in high vorticity regions especially in the shear layer. We attribute this effect to two different causes. First, inertial droplets tend to escape from high vorticity regions which is an established explanation of small-scale clustering~\cite{Toschi.2009}. Second, near high vorticity regions in the shear layer  strong mixing events occur where dry environmental air depleted of droplets is mixed with inner saturated gas laden of droplets. This mixed (non-saturated) regions  lead to a fast droplet evaporation that disappear. The combination of these two phenomena originate a strong clustering.
%
%
This anti-correlation between droplets and vapor phase is better highlighted in figure~\ref{fig:Vor_c} where the instantaneous vapor mass fraction field together with the instantaneous droplet distribution  is provided. It is worth noticing that most part of the droplets are located in the core regions of the turbulent jets in which a strongly inhomogeneous  preferential distribution is observable over the whole downstream evolution of the two turbulent sprays. Nonetheless, in the near-field region, this non-uniform characteristic of droplet dispersion emerges mainly in the mixing layer. This is a source of clustering which propagates also in the downstream evolution. We also emphasize that, in the near-field, a self-preserving core exists with a nearly saturated vapor environment which prevents the droplet to evaporate. %
In both cases, moving downstream the inhomogeneous distribution of the droplets becomes more apparent and can be observed both in the near-axis region and in the jet shear layers. Nonetheless, with the bulk Reynolds number increasing from $Re=6,000$ to $Re=10,000$, the onset of this intensive preferential segregation of the droplets dispersion appears to be slightly shifted in the downstream direction. In addition, we also underline how long is the evaporation length for the higher Reynolds number case where several droplet clusters can be appreciated at a higher distance from the inlet.

\subsection{Eulerian statistics}
\label{sec:eul}
%
%
\begin{figure}[!ht]
    \centering
    \begin{subfigure}[b]{1.\linewidth}
        \centering
        \includegraphics[scale=0.35]{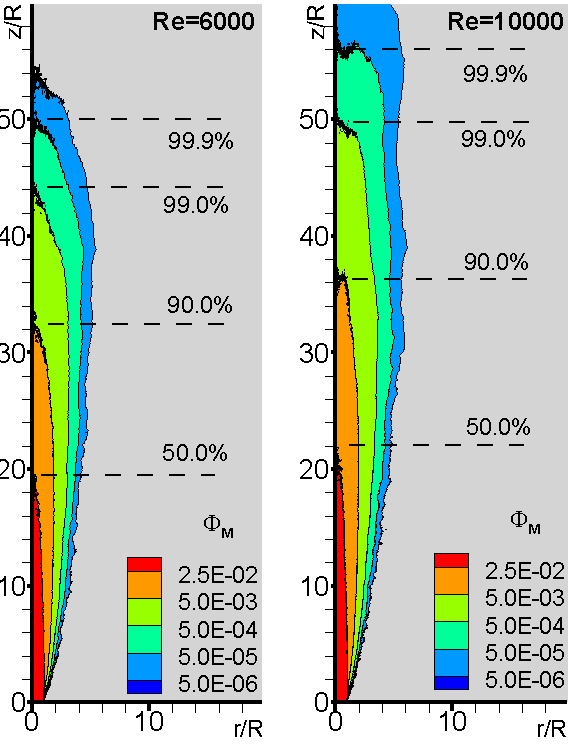}
        \caption{}
        \label{fig:PhiM_a}
    \end{subfigure}
    \begin{subfigure}[b]{1.\linewidth}
        \centering
        \includegraphics[scale=0.35]{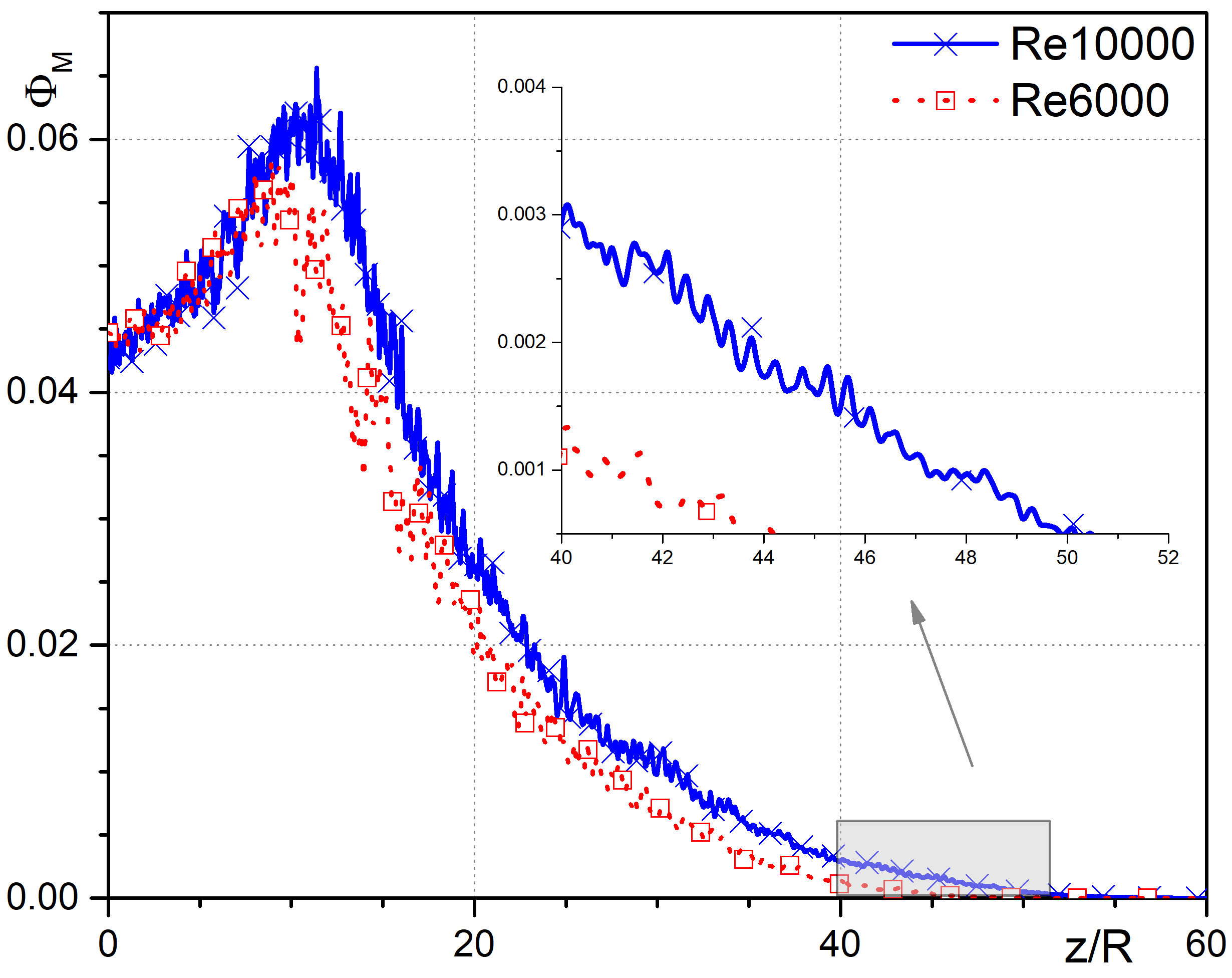}
        \caption{}
        \label{fig:PhiM_b}
    \end{subfigure}
\caption{(a) Averaged Eulerian mass fraction of the liquid phase, $\Phi_M=m_l/m_g$, where $m_l$ and $m_g$ are the mean mass of liquid acetone and air computed inside each mesh cell, respectively. The labels show different distances from the jet inlet section, $z/R$, in correspondence of which the $50\%$, $90\%$, $99\%$, and $99.9\%$ of the overall injected liquid mass is evaporated. (b) Mean liquid mass fraction distribution near the center axis with the range $0<r/R<0.2$.}
\label{fig:PhiM}
\end{figure}
%
\begin{figure}[b]
\centering
    \begin{subfigure}[b]{0.48\linewidth}
        \centering
        \includegraphics[scale=0.35]{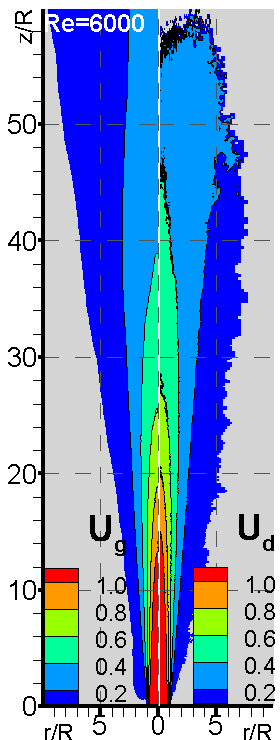}
        \caption{}
        \label{fig:Ugd_a}
    \end{subfigure}
    \begin{subfigure}[b]{0.48\linewidth}
        \centering
        \includegraphics[scale=0.35]{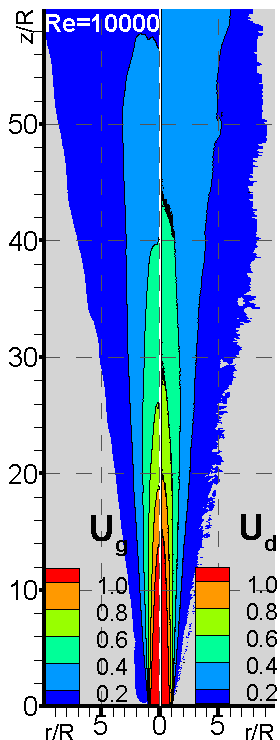}
        \caption{}
        \label{fig:Ugd_b}
    \end{subfigure}
    \begin{subfigure}[b]{1\linewidth}
        \centering
        \includegraphics[scale=0.35]{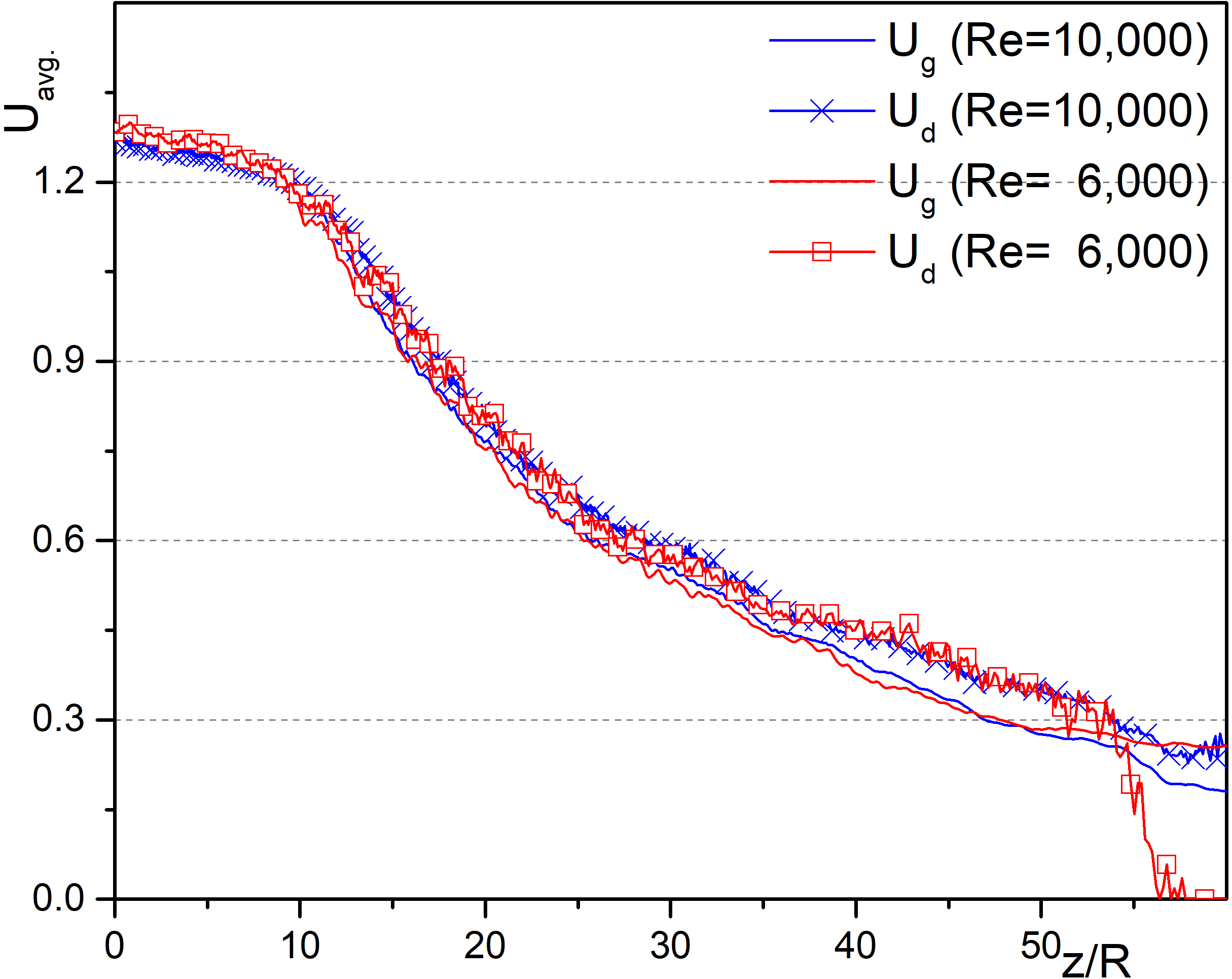}
        \caption{}
        \label{fig:Ugd_c}
    \end{subfigure}
\caption{(a) and (b) show non-dimensional, averaged distribution of gas phase velocity, $U_g$, and dispersed phase velocity, $U_d$, for $Re=6,000$ and $Re=10,000$ turbulent jets, respectively. (c) Mean velocity of two phases near the center axis with the range $0<r/R<0.2$ for both $Re$ jets.}
\label{fig:Ugd}
\end{figure}
%
%
\begin{figure}[]
\centering
    \begin{subfigure}[b]{1.\linewidth}
        \centering
        \includegraphics[scale=0.35]{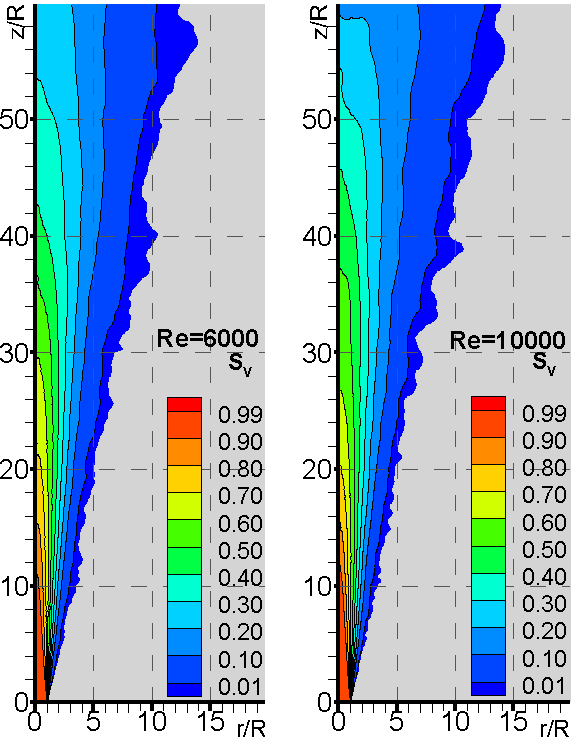}
        \caption{}
        \label{fig:Sv_a}
    \end{subfigure}
    \begin{subfigure}[b]{1.\linewidth}
        \centering
        \includegraphics[scale=0.35]{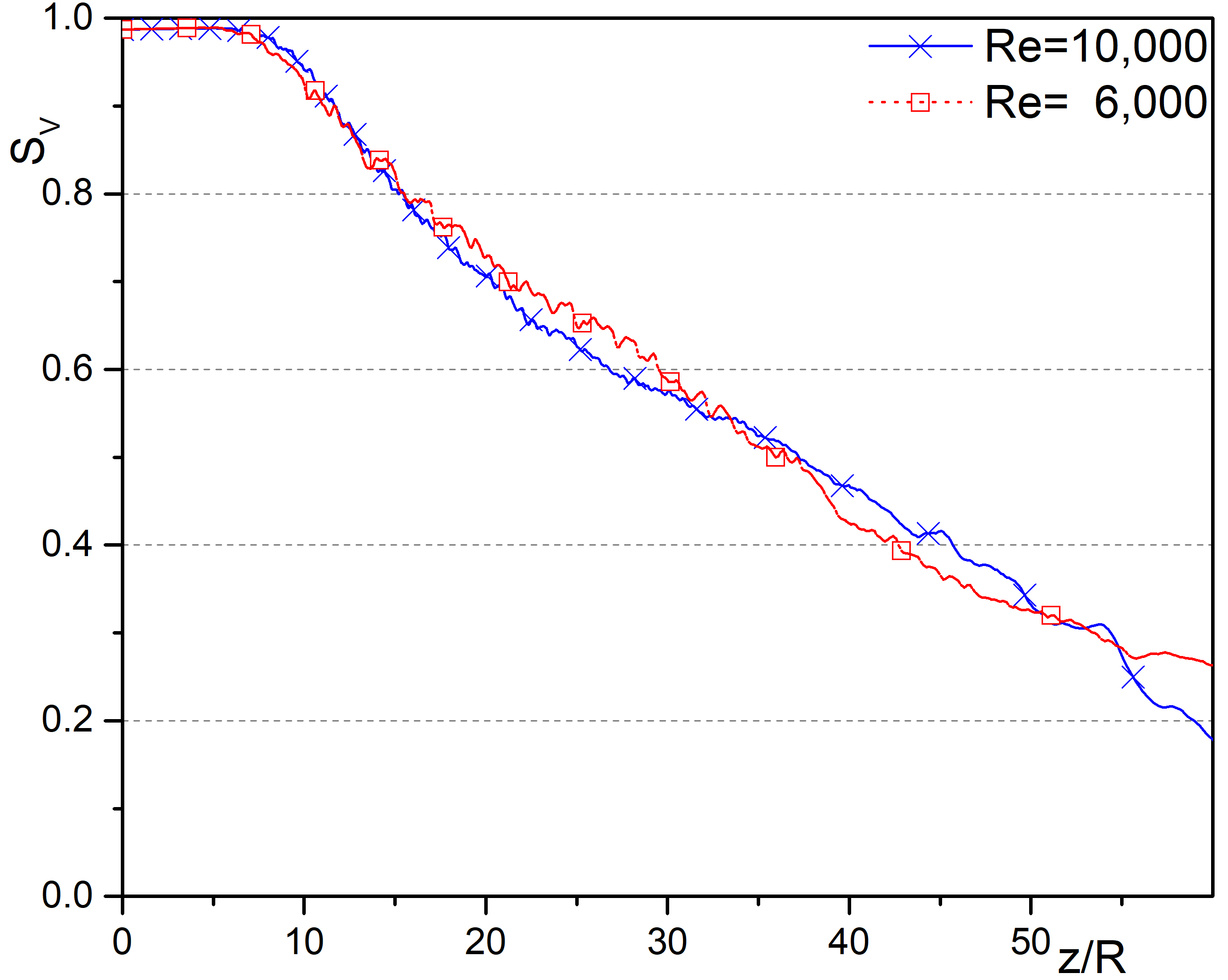}
        \caption{}
        \label{fig:Sv_b}
    \end{subfigure}

\caption{ (a) Averaged saturation field, $S_v = Y_v/Y_{v,s}$, where $Y_v$ is the actual vapor mass fraction and $Y_{v,s} = Y_{v,s}(p_0,T)$ is the value of vapor mass fraction evaluated at local saturation condition. (b) Mean vapor saturation distribution near the center axis with the range $0<r/R<0.2$.}
\label{fig:Sv}
\end{figure}
%
Figure~\ref{fig:PhiM}(a) provides the distribution of the average liquid mass fraction, $\Phi_M$. The mass fraction is defined as $\Phi_M= m_l / m_g$, where $m_l$ and $m_g$ are the mass of the liquid acetone and the mass of gaseous phase evaluated inside each mesh cell. 
A spray vaporization length is defined as the axial distance from the inflow section to where the droplets lost, in average, $99\%$ of their initial mass~(\citet{DallaBarba.2018}). According to this definition, the vaporization process ends at about $z/R \simeq 44$ and $z/R \simeq 50$ for $Re=6,000$ and $Re=10,000$, respectively. The vaporization length of the high-Reynolds jet is about $13\%$ longer than the low-Reynolds one.
An additional comparison of the spatial distribution of the average liquid mass fraction between the two cases is provided in figure~\ref{fig:PhiM_b} that shows the centerline mean liquid mass fraction versus the axial coordinate, $z$. 
The centerline behavior exhibits a near-field hump: the liquid mass fraction increases along the jet axis from the inlet section up to a peak located at $z/R \simeq 9$ for $Re=6,000$ and $z/R \simeq 11$ for $Re=10,000$. Further downstream, $\Phi_M$ reduces along the jet axis until droplets inside the turbulent spray totally evaporate. The centerline behavior of $\Phi_M$ keeps higher values in the $Re=10,000$ case than in its counterpart. 
This behavior is consistent with the  observations by~\citep{Picano.2010,Lau.2014,Lau.2016} in which the centerline concentration of inertial particles is found to increase above the 
exit value. The phenomenon can be explained considering the interplay of the droplet/particle inertia and the decay of the mean fluid  velocity. The inertia induces a delay of the particle velocity to adapt to the slower (decayed) flow velocity creating a local concentration peak, see~\citet{Picano.2010} for more details. 
\par 
Fluid and particle mean velocities are displayed in figure~\ref{fig:Ugd}. In the panels \ref{fig:Ugd_a} and \ref{fig:Ugd_b} for each case is shown the mean gas velocity in the half-left image and the mean droplet velocity in the half-right image. As apparent, the dispersed phase velocity is slightly higher than the corresponding fluid velocity, but no significant differences emerge comparing the jets at the two Reynolds numbers. The centerline values of the same quantities, shown in panel \ref{fig:Ugd_c}, confirm this overall impression. Given that, the macroscopic flow behaviors of present cases is not affected by the Reynolds number, so the difference found in the evaporation lengths is not induced by a deviation of the gas and particle velocity fields.  
\par
\begin{figure}[!htb]
    \centering
    \begin{subfigure}[b]{0.475\linewidth}
        \centering
        \includegraphics[width=1.\linewidth]{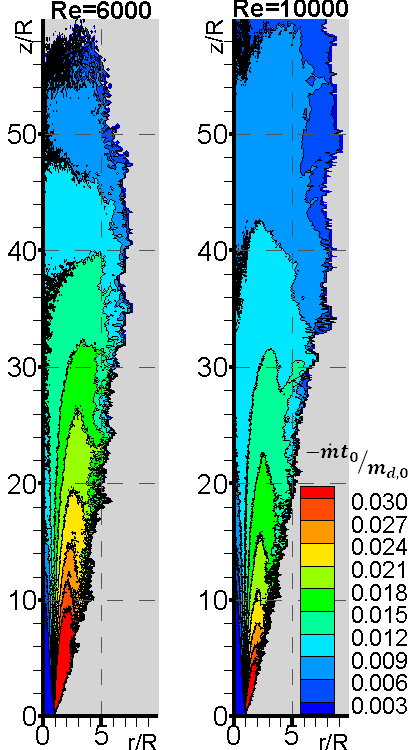}
        \caption{}
        \label{fig:evp_a}
    \end{subfigure}
    \begin{subfigure}[b]{0.475\linewidth}
        \centering
        \includegraphics[width=1.\linewidth]{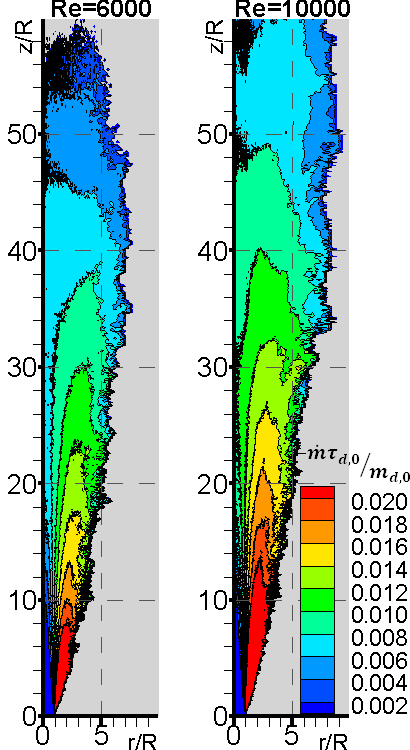}
        \caption{}
        \label{fig:evp_b}
    \end{subfigure}
\caption{(a) Average droplet vaporization rate, $\langle \dot m_d\rangle$ divided by the reference mass-flow-rate scale $\dot{m}_{d,0} = m_{d,0}/t_0$ with $m_{d,0}$ the initial droplet mass and $t_0$ the reference time scale of the jet. (b) is similar to (a) except for the normalization time scale, $\tau_{d,0}$.}
\label{fig:evp}
\end{figure}

In a turbulent spray, as the turbulent core spreads and slowly decays, the dry and irrotational environmental air surrounding the jet is entrained continuously diluting the vapor concentration and permitting the overall vaporization process to advance. Since the inner core fluid does not reach the outer jet dry region, but is just diluted by the entrainment, the centerline region presents a higher saturation level over the whole downstream evolution of the flow. 
Mean saturation field, $S_v$, is provided in figure~\ref{fig:Sv_a}. Both  flows are almost saturated near the inlet section, as prescribed by the inlet conditions. The saturation level gradually decreases in the downstream direction, maintaining a sharp gradient towards the outer jet region. No significant differences are found between the two cases.  Some small discrepancies are present in the intermediate and far-field region, $z/R\ >\ 20$. The case at $Re=6,000$ (left panel) shows a relatively higher saturation level in the region $20< z/R <30$ while the opposite occur beyond $z/R \simeq 30$. 
As will be further addressed in the following, at lower Reynolds number we observe a faster evaporation rate in the first part of the spray, which explains the higher saturation level in this region. Conversely, in the far-field there are more  dispersed evaporating droplets at the higher Reynolds number, so we expect a higher saturation level.   
%
%
\begin{figure}[]
    \centering
        \begin{subfigure}[b]{1.\linewidth}
        \centering
        \includegraphics[scale=0.35]{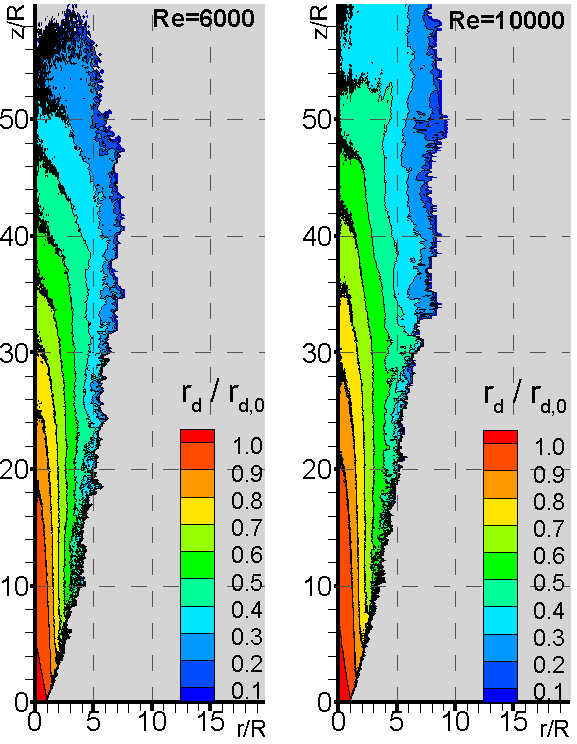}
        \caption{}
        \label{fig:rd_a}
    \end{subfigure}
    \begin{subfigure}[b]{1.\linewidth}
        \centering
        \includegraphics[scale=0.35]{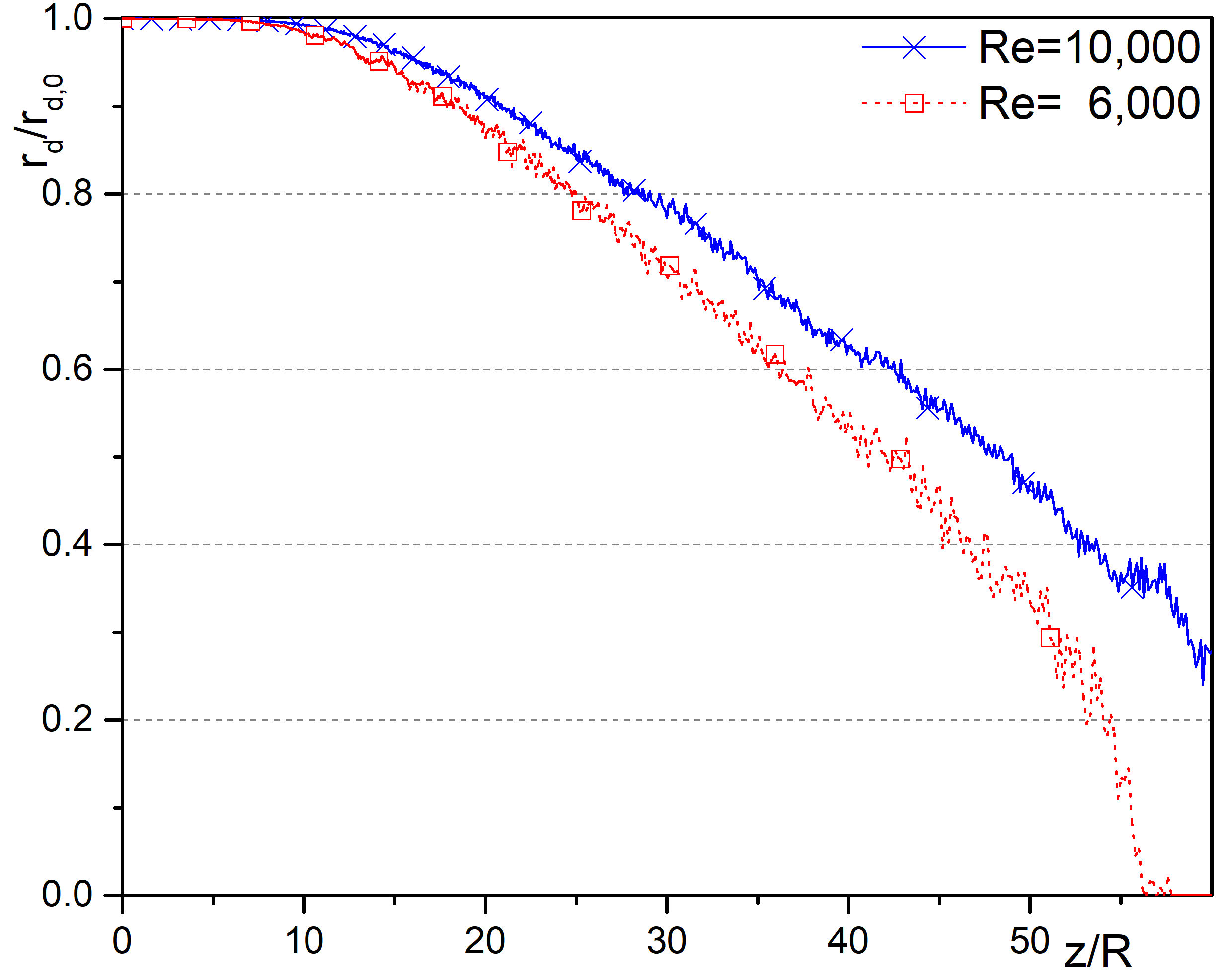}
        \caption{}
        \label{fig:rd_b}
    \end{subfigure}
\caption{(a) Average droplet radius, $\langle r_d\rangle$, scaled by the droplet initial radius $r_{d,0} = 6 \mu m$. (b) Mean droplet radius distribution near the center axis with the range $0<r/R<0.2$}
\label{fig:rd}
\end{figure}
%
Figure~\ref{fig:rd} provides a comparison between the two cases for the mean droplet radius, $\langle r_d\rangle$. %
In both cases, in correspondence of each axial position, $z/R$, larger droplets are always located in the jet core region whereas smaller droplets can be found towards the mixing layer. This behavior is expected since the saturated jet core reduces the vaporization rate of the droplets near the  centerline. On the other hand, the entrainment of environmental dry air dilutes the acetone vapor concentration in the jet mixing layer enhancing the vaporization process.
At the higher Reynolds number, we find larger droplets denoting a slower average evaporation rate. A more quantitative view assessing this aspect is provided  in panel \ref{fig:rd_b} where the centerline value of the mean droplet radius is shown.
The mean evaporation rate is shown in figure~\ref{fig:evp_a} where the droplet evaporation rate is normalized by the flow time scale $t_0$ and the initial droplet mass $m_{d,0}$. 
Evaporation peaks in the mixing layer and is more intense in the near field. The observed behavior is consistent with the mean droplet radius previously discussed. 
From quantitative point of view, the case at $Re=10,000$ shows less intense values which explain the longer evaporation length. Dealing with dimension-less quantities, droplet mass transfer is proportional to $1/St \propto 1/Re$, implying a relatively slower mass transport with respect to advection time scale. Nonetheless,
the dissipative length-scale, as well as the finer length-scales of turbulent mixing-layer  structures, become smaller in the high-Reynolds case, leading to a faster time-scale of the entrainment process arising in the mixing layer.  This peculiar  feature makes the (non-dimensional) entrainment, spreading and decay rates of turbulent jets independent of the Reynolds number~\cite{Pope.2001}. Being the dispersed droplets inertial, they are not able to get fully benefit of these \emph{fast} features and globally show a slower evaporation rate. To better characterize this aspect, in figure~\ref{fig:evp_b} we show the mean droplet evaporation rate normalized by the initial droplet relaxation time $\tau_0=St\,t_0$ and the initial droplet mass. This normalization takes into account the droplet inertia and would remove the Reynolds number direct dependence. Hence, we observe that at higher Reynolds number the mean ($\tau_0$-dimensionless) evaporation rate appears a bit higher with respect to the case at $Re=6,000$. The more intense fluctuations of the higher Re-number case fasten the evaporation rate normalized with the droplet-relaxation time, i.e.\ accounting for the proper droplet time-scale. However, from physical point of view, this small growth is not sufficient for compensating the faster advection time-scale $t_0$ induced by the higher inflow velocity of $Re=10,000$. Hence, this explains the longer jet evaporation rate which, nevertheless, does not scale proportionally to the Reynolds number. %

%
\par
\par
As previously stated, a strong preferential segregation of the dispersed phase rises, both in the low and high Reynolds number cases. %
Several mechanisms governing the preferential concentration of droplets in a turbulent flow have  been proposed, e.g.\ the small-scale clustering~\citep{Toschi.2009}, the sweep-stick mechanism~\citep{Goto.2008}, the accumulation of droplets along jet axis~\citep{Picano.2010,Lau.2014,Lau.2016} and the intermittent dynamics of the jet mixing layer~\citep{DallaBarba.2018}. 
Independently from the phenomena giving rise to the observed inhomogeneous and preferentially-segregated spatial distribution of droplets, within clusters the local vapor concentration is significantly higher than its bulk counterpart. Consequently, the evaporation process of droplets located into a cluster is significantly slowed down. The vaporization may be even locally stopped, if the vapor concentration reaches the saturation level, $Y_{v,s}(p_0,T)$, producing a non-evaporating core around the cluster~\citep{Reveillon.2007}.
In literature, several approaches have been proposed to measure the intensity of the preferential segregation of the dispersed phase in droplet or particle laden multiphase flows~\citep{Shaw.2002}. Here, in order to provide a spatial map we employ the  clustering index~\citep{Battista.2011,DallaBarba.2018}:
\begin{equation}
\label{eq:cl}
K = \frac{\overline{(\delta n)^2}}{\overline{n}} - 1.
\end{equation}
To compute $K$ in equation above,\eqref{eq:cl}, the Eulerian domain is discretized employing an uniform and equispaced Cartesian grid of cubic sampling cells. The edge size of the cells is set to $L/R\ =\ 0.2$. The variables $\overline{n}$ and $\overline{(\delta n)^2}$ are the mean and the variance of the number of droplets located into each sampling cell, respectively. The clustering index, $K$, is vanishing for any cell where the distribution of the droplets is random (Poisson process), whereas $K$ becomes positive when the variance exceeds the mean value due to the existence of a preferential concentration of droplets within the considered cell. In this sense, large positive values of $K$ correspond to a strong segregation of the dispersed phase whereas low values to a uniform spatial distribution of droplets. The results are presented in figure~\ref{fig:K_St_a}, showing the contour plot of the mean clustering index computed by averaging the instantaneous and local values of $K$ along the azimuthal direction and over time. The distribution and the intensity of the preferential segregation of the dispersed phase present only a weak dependence from the bulk Reynolds number of the jet, since no strong differences exist in the spatial distribution of $K$, at least for the cases considered in this paper. In the immediate proximity of the inflow section, $K$ assumes positive values mainly in the mixing layer, whereas it is only weakly positive in the core. On the other hand, in the intermediate and far-field, $K$ assumes quite large positive values over the whole turbulent jet, both in the core and mixing layer. The peak values for the clustering index  occur in the jet core, between $z/R\simeq10$ and $z/R\simeq20$ in both low and  high Reynolds cases, the latter presenting a small shift in the downstream direction of the $K$-peaking-region.
\par
To determine the leading mechanism driving the preferential segregation of droplets in different spray regions, we consider the droplet Stokes number evaluated at the dissipative time scale. Figure~\ref{fig:K_St_b} provides the trend of the mean Stokes number $St_\eta$, defined as the ratio of the droplet response time, $\tau_d$, and the characteristic time of the dissipative scale, $\tau_{\eta} = (\nu / \epsilon)^{1/2}$, computed along the jet axis and plotted versus $z/R$. The local dissipative time scale is evaluated in correspondence of each cell of the computational Eulerian grid, located within the radial range $0<r/R<0.2$. The computed time scale is then adopted to estimate the Stokes number of the droplets located in the correspondent cell. Finally, the mean Stokes number is computed by averaging its instantaneous and local values along the azimuthal direction, within the considered radial range as well as over time, concurrently. The axial trend of $St_\eta$ shows a distinct hump around $z/R\simeq 15$, which is followed by a gradually decrease along the streamwise direction until a unity value of is achieved around $z/R \simeq 30$ and $z/R \simeq 40$, at $Re=6,000$ and $Re=1000$, respectively. The peak values for the Stokes number are achieved in correspondence of the peak of the clustering index. Small-scale inertial clustering rises from the competition between the droplet inertia and the Stokes drag. The drag tends to force the droplets to follow the highly convoluted paths of the turbulent motion, whereas the finite inertia of the droplets prevents them to exactly move along these material-paths. By this mechanism, largest droplets, for which $St_{\eta} \gg 1$, act as ballistic particles with respect to the smallest scales of turbulence. These droplets move across the smallest turbulent structures being only weakly perturbed and do not tend to accumulate in clusters. On the contrary, the smallest droplets, for which $St_{\eta} \ll 1$, act as passive tracers which follow exactly the path of the local turbulent motion. Also these droplets do not contribute significantly to inertial clustering. Finally, droplets for which $St_{\eta} \simeq 1$ manifest an intermediate dynamics, accumulating into the small-scale interstitial vortical regions and giving rise to strong inhomogeneities of the spatial distribution of the dispersed phase. Based on the trend of $St_{\eta}$, since the Stokes number keeps values significantly higher than unity in the intermediate field, we assess that inertial clustering must play a minor role in the preferential segregation of the liquid phase in this region, whereas it becomes dominant in the far field of the jet being not-negligible in the near-field, as well. Different mechanism, rather than small-scale inertial clustering, should be considered to explain the clustering of droplets in the near and intermediate field of the jet. We attribute the rising of the preferential concentration of droplets in these regions to the effect of the intermittent dynamics of the mixing layer of the jet. In the outer-spray region, high-intensity vortical structures separate the turbulent core, populated by droplets, by the  dry environment, depleted of droplets. By an Eulerian point of view, this region is characterized by large temporal and spatial fluctuations of the local droplet concentration because of the intermittent dynamics of the turbulent-non-turbulent interface. The entrainment causes the envelopment of dry-air bubbles, depleted of droplets, within the turbulent jet. %
These dry and droplet-depleted bubbles mix with high-saturated gaseous regions full of droplets, giving rise, instant by instant, to large inhomogeneities of the droplet distributions which are, then, advected downstream. %
To highlight this source of clustering in the mixing layer we show in figure~\ref{fig:K_St_c} the mean vapour  mass fraction variance $\langle {Y'}^2_v \rangle$ for the two cases. It is possible to note that the location of the peak of $K$ and $\langle {Y'}^2_v \rangle$ 
in the near field region is similar. 
Hence we believe that the initial source of droplet clustering is induced by the intermittent dynamics of the mixing layer which create clusters that are advected downstream. Then in the intermediate and far-field regions inertial clustering becomes more important.  
%
%
\begin{figure}[!b]
    \centering
    \begin{subfigure}[b]{0.99\linewidth}
        \centering
        \includegraphics[scale=0.35]{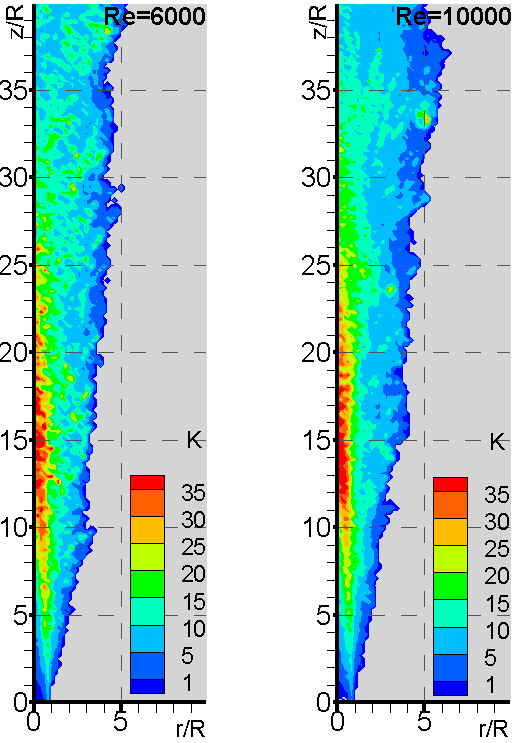}
        \caption{}
        \label{fig:K_St_a}
    \end{subfigure}
    \begin{subfigure}[b]{0.99\linewidth}
        \centering
        \includegraphics[scale=0.35]{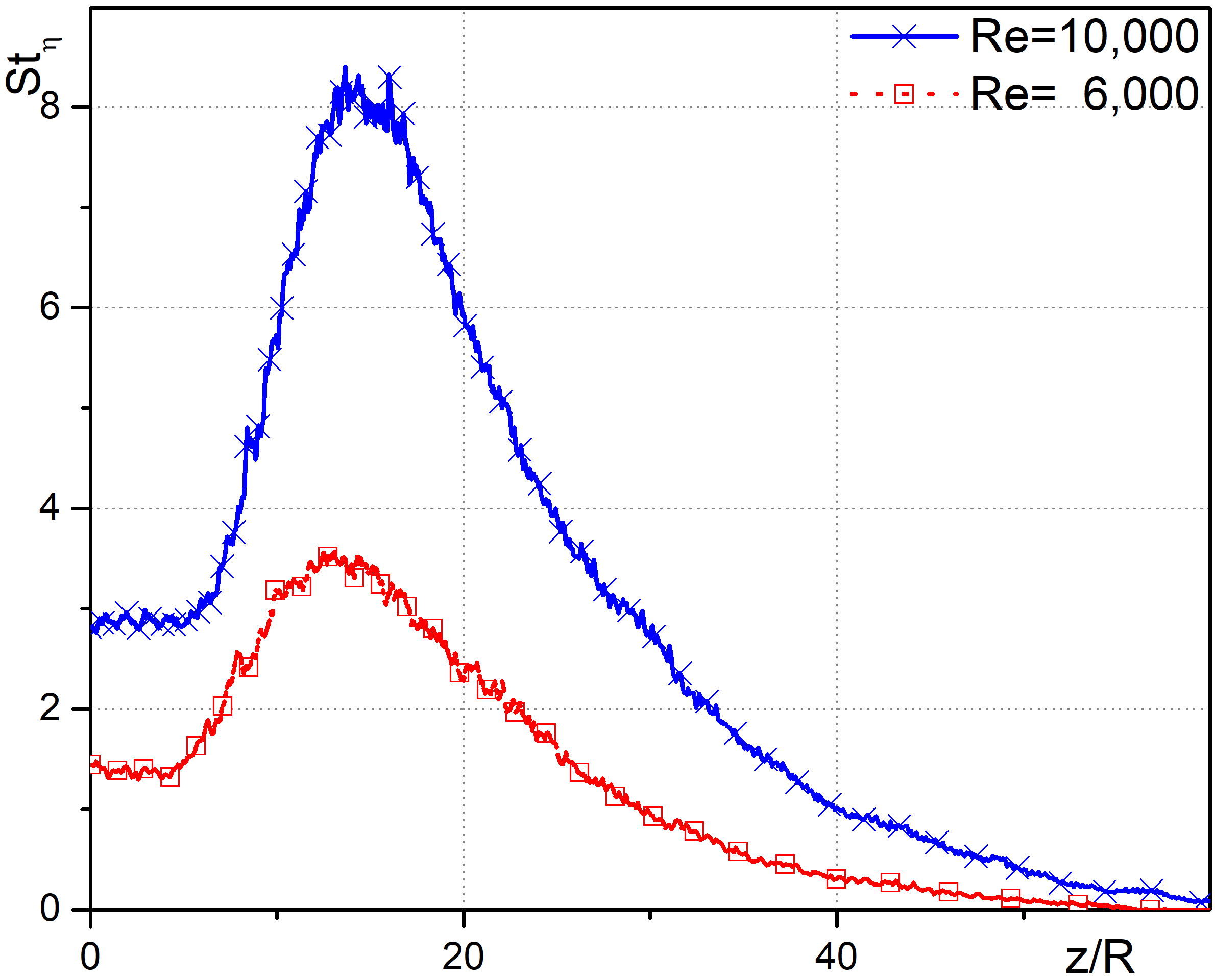}
        \caption{}
        \label{fig:K_St_b}
    \end{subfigure}
\end{figure}
\begin{figure}[ht]\ContinuedFloat
    \centering
    \begin{subfigure}[b]{0.99\linewidth}
        \centering
        \includegraphics[scale=0.35]{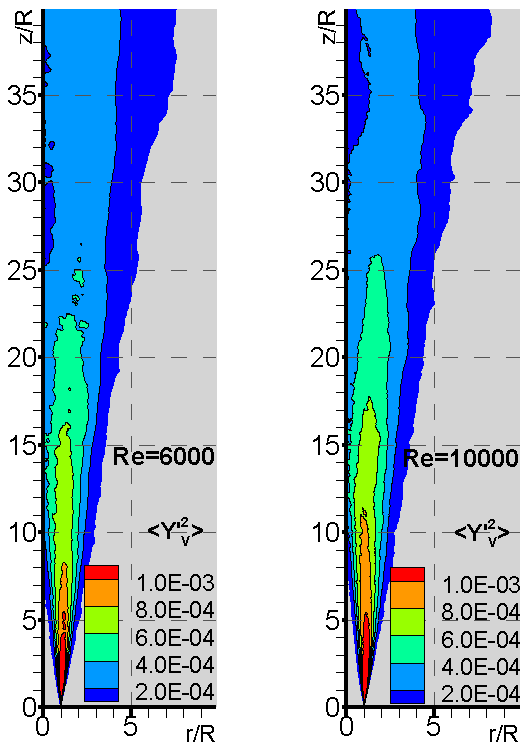}
        \caption{}
        \label{fig:K_St_c}
    \end{subfigure}
\caption{(a) Contour plot of the mean droplet clustering index, $K$, defined according to equation~\ref{eq:cl}.  (b) Evolution along the jet axis of the mean droplet Stokes number based on the dissipative time-scale of the carrier phase, $St_{\eta} = \tau_d/ \tau_{\eta}$. (c) Average distribution of variance of vapor mass fraction,$\langle Y'^2_v \rangle$.}
\label{fig:K_St}
\end{figure}
%
\par
\begin{figure}[]
    \centering
    \begin{subfigure}[b]{1.\linewidth}
        \centering
        \includegraphics[scale=0.35]{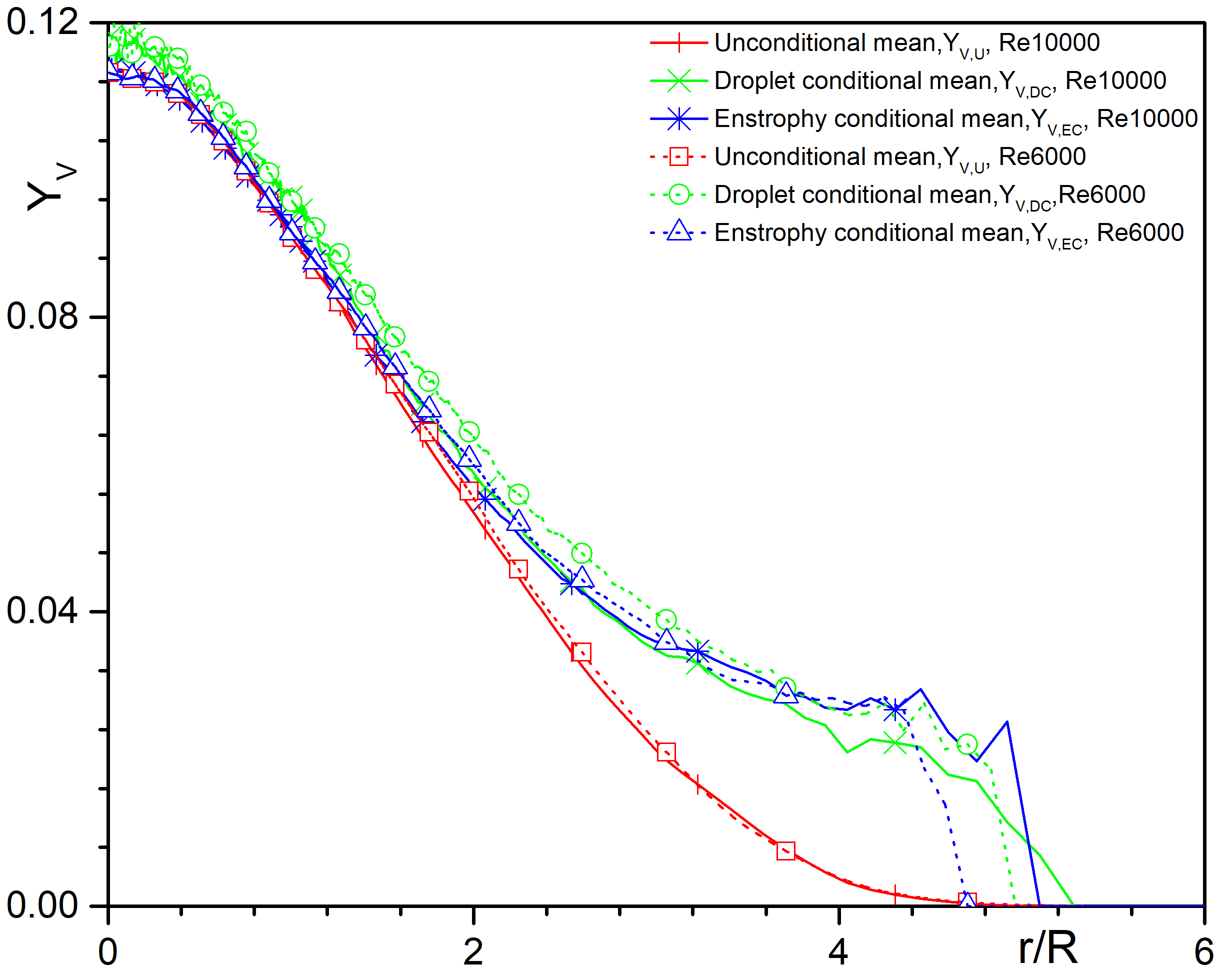}
        \caption{}
        \label{fig:Yv_a}
    \end{subfigure}
    \hfill
    \begin{subfigure}[b]{1.\linewidth}
        \centering
        \includegraphics[scale=0.35]{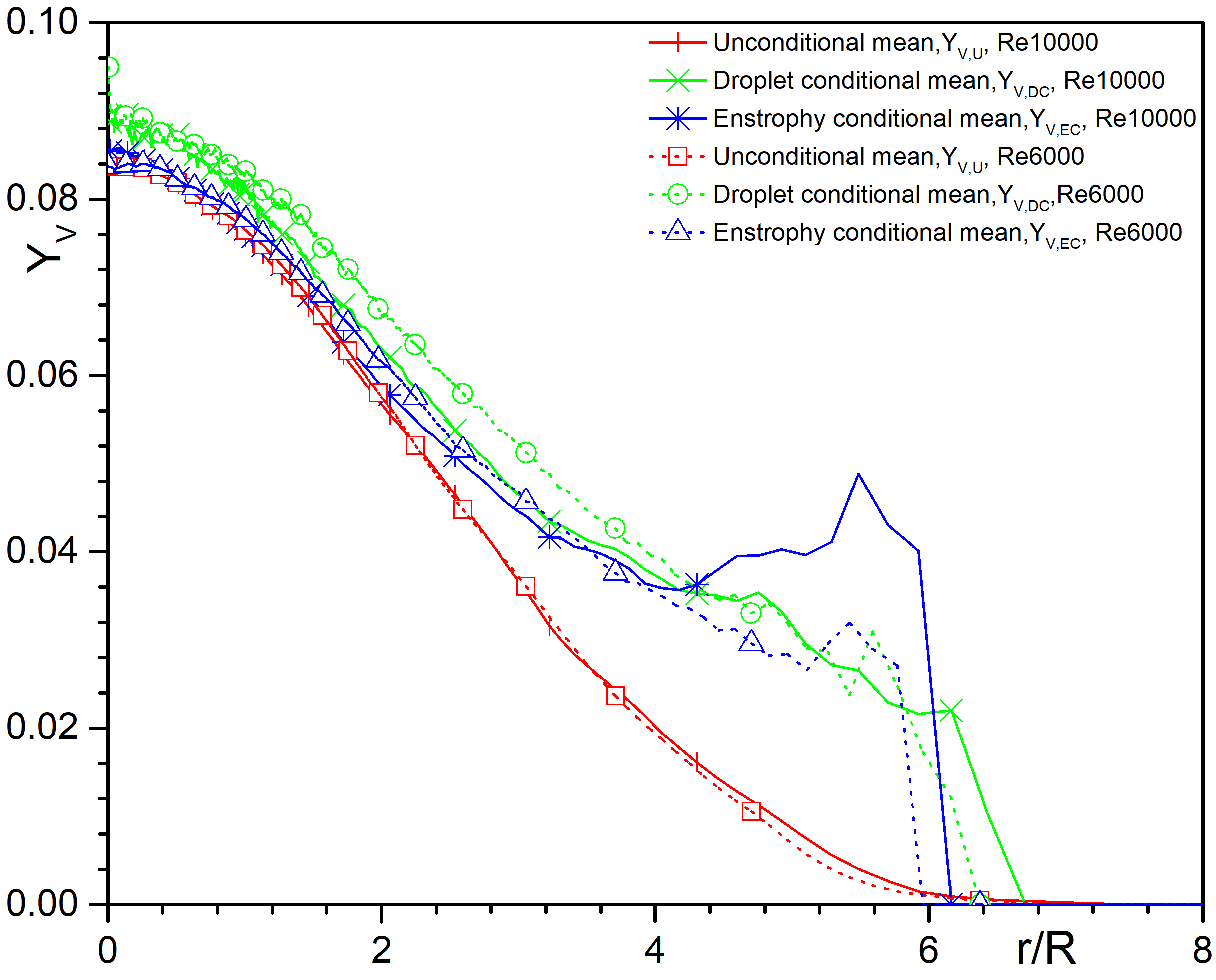}
        \caption{}
        \label{fig:Yv_b}
    \end{subfigure}
\caption{Radial profiles of the average vapor mass fraction field at four different axial distances from the origin: (a) $z/R = 20$, (b) $z/R = 30$. Each plot shows the enstrophy-threshold conditional average, $Y_{V,EC}$, the droplet-presence conditional average, $Y_{V,DC}$,and the unconditional Eulerian one, $Y_{V,U}$. The enstrophy-threshold conditional average is calculated by sampling the vapor mass fraction only over turbulent core events ($I = 1$), that is, when local enstrophy exceeds a fixed threshold. $Y_{V,DC}$ is the vapor concentration field obtained by a conditional average on the droplet presence in a given point.}
\label{fig:Yv}
\end{figure}

To quantify the importance of the droplet preferential segregation on the evaporation process, we provide a comparison between the conditional mean vapor concentration sampled by the droplets (\textit{droplet-conditional}), $Y_{V,DC}$, and the unconditional Eulerian one, $Y_{V,U}$. The latter is computed by averaging the vapor mass fraction field along the azimuthal direction and over time, concurrently. The former is defined as the average of the Eulerian vapor mass fraction field, conditioned to the presence of a droplet at a given point. The statistics is computed on the Eulerian grid, associating to each grid cell, the mean vapor mass fraction sampled by the droplets instantaneously located within the considered cell.  
Figure~\ref{fig:Yv} shows the radial profiles of $Y_{V,DC}$ and $Y_{V,U}$ at two different axial distances from the jet inflow, $z/R=20$ and $z/R = 30$, respectively. Globally, the vapor mass fraction sampled by  droplets results to be higher than the correspondent unconditional value, regardless of the Reynolds number. As already reported in~\citep{DallaBarba.2018}, this preferential sampling of the vapor mass fraction field actuated by the droplets can be related to different mechanisms. The primary contribution originates from the vapor self-produced by   droplets with a tracer behavior ($St_{\eta}\ll 1$). These droplets exactly follow the path of the turbulent motion of the surrounding vapor atmosphere, increasing the local vapor concentration in their surrounding material volume, and hence sampling higher vapor concentration. Secondarily, within a droplet cluster, independently from the mechanisms driving its formations, a highly-saturated cloud establishes in its proximity. Hence, in the presence of preferential segregation, an oversampling of the vapor concentration field is expected also by the droplets evaporating inside clusters.
%
%
%
%
The radial profiles of the mean vapor concentration are similar  inside the jet core,~Fig.\ref{fig:Yv_a}, whereas significant differences appear in the mixing layer. In the core region the jet presents a nearly uniform vapor mass fraction field that nullifies the oversampling actuated by the droplets. On the other hand, in the mixing layer, the entrainment of bubble of dry air depleted of droplets causes the Eulerian, unconditional statistics to assume lower values than the conditional one. This aspect is strictly connected to the clustering mechanism previously discussed. %
Further downstream, the difference between the \textit{droplet-conditional} vapor concentration and the unconditional one gradually increases, Fig.\ref{fig:Yv_b}, even in the jet core. %
In the far-field, the inertial small-scale clustering plays a dominant role in the preferential segregation of droplets as shown in figure~\ref{fig:K_St}. %
This lead to a considerable increment of the vapor mass fraction field sampled by the droplets grouped in clusters with respect to the unconditional statistics. Besides, no evident disagreement is observable for the unconditional mean vapor concentration profiles in $Re=10,000$ and $Re=6,000$ cases. 
\par
To better analyze the mixing layer of jet spray, we remind that the inner core and the irrotational outer region are separated by a nearly-sharp fluctuating interface, which is highly convoluted over a wide range of vortical scales~\citep{daSilva.2014}. Different approaches can be employed to identify this interface~\citep{Bisset.2002,Westerweel.2005,Holzner.2007,Jimenez.2010}. Here, we consider the local enstrophy,  $\zeta^2\ =\ \Vert \nabla \times \overrightarrow{u} \Vert$. The inner turbulent core is characterized by large fluctuation of enstrophy, while in the outer region enstrophy is null. Thus, fixing a threshold, $\zeta^2_{th}$, it is possible to distinguish if a given Eulerian position at fixed time is located in the turbulent jet core or in the outer irrotational environment employing the following index:
\begin{equation}
I(\boldsymbol{x},t) = H[\zeta^2(\boldsymbol{x},t)-\zeta^2_{th}],
\end{equation}
with $H$ the Heaviside function. $I\ =\ 1$ denotes a turbulent event at Eulerian position $\boldsymbol x$, while $I\ =\ 0$ an irrotational one. The conditionally average profile of vorticity magnitude has been proven to weakly depend on the threshold value in the range $\zeta_{th} \approx 0.7U_0 / \delta$, with $U_0$ the mean velocity scale of the wake and $\delta$ the shear layer thickness~\citep{Bisset.2002,daSilva.2014}. Hence, the $I$ index, employed here to distinguish between the rotational jet core and irrotational environment, is weakly dependent from the threshold value $\zeta_{th}$. In~\citet{DallaBarba.2018}, it has been observed no relevant differences on the $I$ by changing the value of $\zeta_{th}\ =\ 0.6 U_0 / R$ by a factor 2. Consistently, the same threshold value is adopted in the present work. An enstrophy conditional average for the vapor mass fraction field, $Y_{V,EC}$, can be defined as the mean vapor mass fraction field obtained by only sampling the instantaneous values of $Y_V(\boldsymbol{x},t)$ linked to turbulent events, that is only for time and positions where $I=1$. Figure~\ref{fig:Yv} provides, together $Y_{V,DC}$ and  $Y_{V,U}$, the radial profile of the enstrophy conditioned $Y_{V,EC}$. The unconditional and both the conditional statistics upstream $z/R \simeq 20$ are similar in the core, whereas in the mixing layer the radial profile of $Y_{V,EC}$ is similar to that of $Y_{V,DC}$, both the statistics assuming higher values than that of $Y_{V,U}$. This difference, confined to the outer spray regions, confirms the contribution of the intermittent dynamics of the mixing layer to the non-homogeneous spatial distribution of the dispersed phase discussed above. Droplets moving towards the outer region are enclosed by a high concentration vapor cloud which is expelled from the turbulent jet core. Simultaneously, dry air without droplets is engulfed in the core enhancing the fluctuation of droplets distribution and vapor concentration. Further downstream, $z/R \simeq 30$, the droplet-conditional vapor concentration profile shows the same shape as the enstrophy-conditional one, except for an almost constant offset, both in the core and mixing layer. This confirms how, in the far field, both the small-scale inertial clustering and the intermittent dynamics of the mixing layer contribute to the oversampling of the vapor mass fraction field actuated by the dispersed phase, the former mechanism becoming the dominant effect in the jet core. The combination of these two mechanisms, leading to the observed oversampling, strongly reduces the vaporization rate with respect to the one that would be estimated employing unconditioned statistics. The phenomenologies described above are similar for both the cases considered in this study. In particular, no significant difference is observable for unconditional vapor concentration profiles between $Re=6,000$ and $Re=10,000$. 

\subsection{Lagrangian statistics}
\label{sec:lag}
Turbulent fluctuations are responsible for extremely different Lagrangian histories of the evaporating droplets. 
%
%
\begin{figure}[]
\centering
    \begin{subfigure}[b]{1.\linewidth}
        \centering
        \includegraphics[scale=0.032]{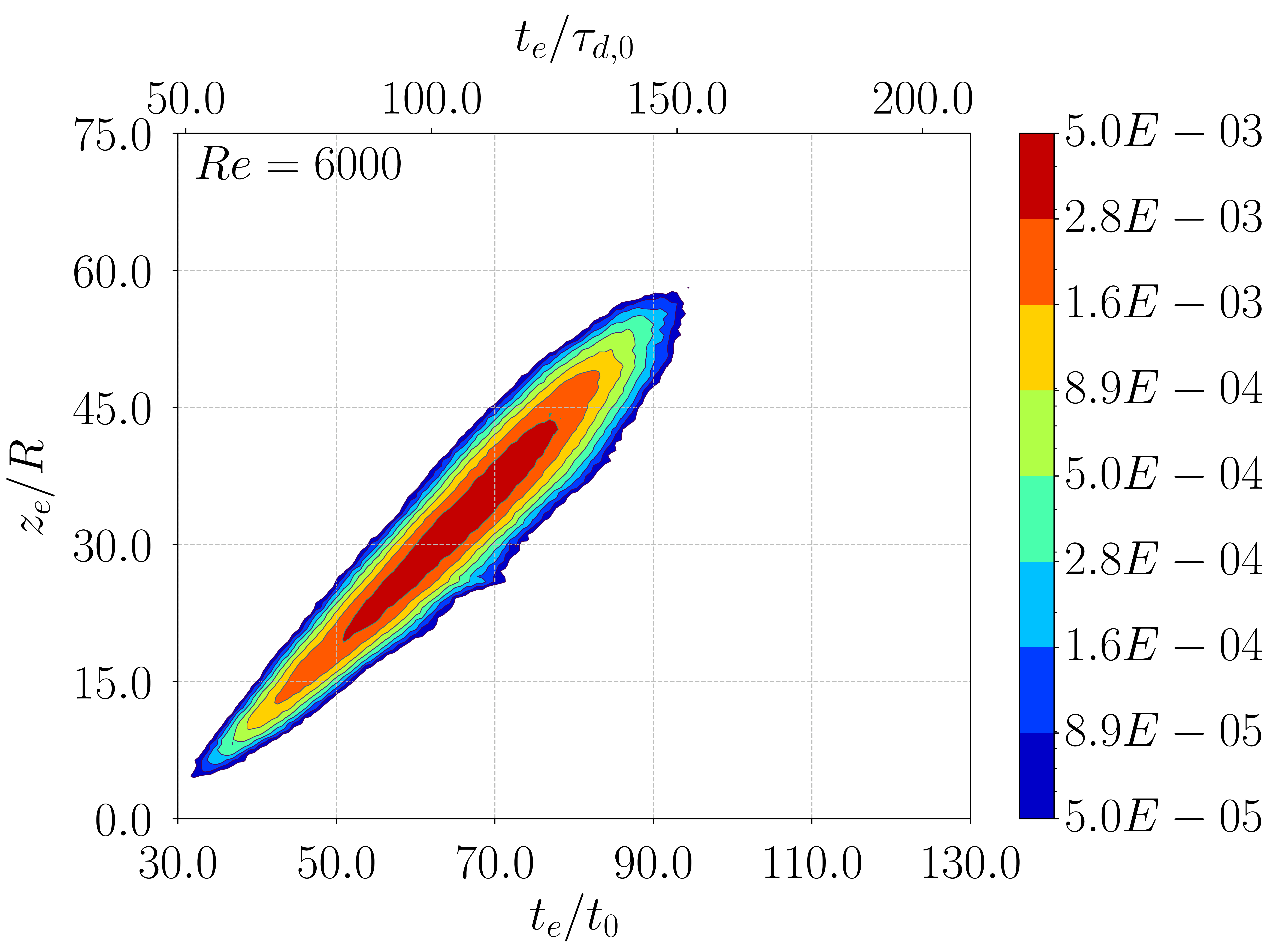}
        \caption{}
        \label{fig:JPDF_Zdt_a}
    \end{subfigure}
    \begin{subfigure}[b]{1.\linewidth}
        \centering
        \includegraphics[scale=0.032]{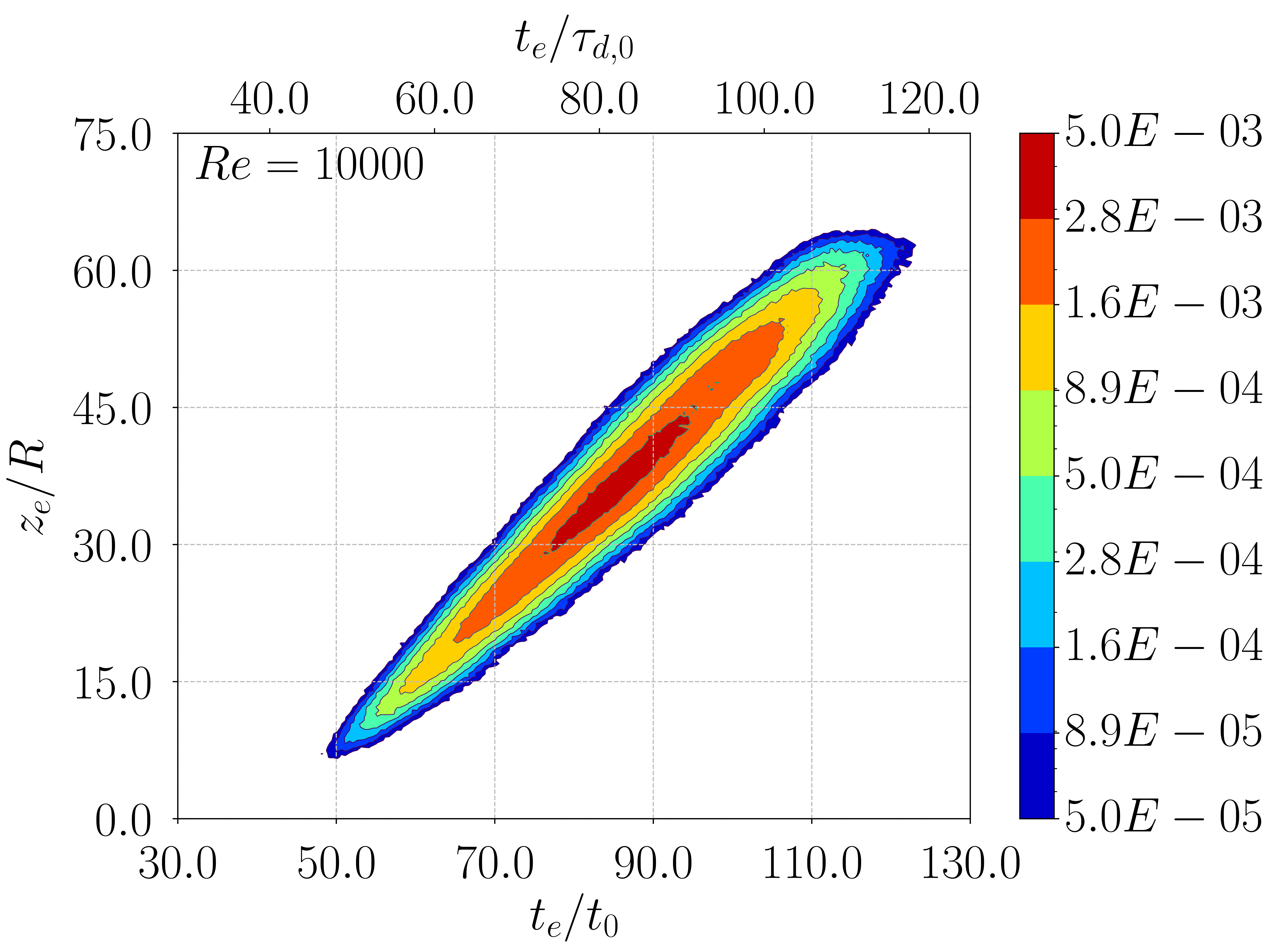}
        \caption{}
        \label{fig:JPDF_Zdt_b}
    \end{subfigure}
\caption{JPDF of the vaporization lengths and times of the injected droplets computed over the whole droplet population for $Re = 6,000$ (a) and $Re = 10,000$ (b) cases. The lower axis provides the non-dimensional time, $t_e/t_0$, whereas the secondary top axis provides the non-dimensional time, $t_e/\tau_{d,0}$.}
\label{fig:JPDF_Zdt}
\end{figure}
The joint probability density function (JPDF) of the droplet vaporization lengths and time are reported for both cases in figure~\ref{fig:JPDF_Zdt}. The vaporization length for a single droplet is defined as the axial distance from the inlet, $Z_e/R$, where the droplet radius is decreased from $r_{d,0}$ to a threshold radius, $r_{d,th} = 0.1 r_{d,0}$ (99.9\% of the mass evaporated). The correspondent time is defined as the droplet vaporization time, $t_e / t_0$. As it can be appreciated from the figure, for both $Re = 6,000$ and $Re = 10,000$ cases, there is a strong linear correlation between the droplet vaporization length and time, whereas different slope rates are observable. It is also remarkable how different the droplets histories are: half of ejected droplets are still present around $z_e/R \simeq 32$ for $Re = 6,000$ and $z_e/R \simeq 36$ for $Re = 10,000$, where 90\% of the droplet mass is evaporated (see figure \ref{fig:PhiM}). Correspondingly, the median vaporization times are about $t_e/t_0 = 65$ and $t_e/t_0 = 90$ for $Re=6,000$ and $Re=10,000$, respectively. %
Consistently with previous discussions, we note a slower average vaporization rate for the higher Reynolds number. %
In addition we note that the actual droplet evaporation time varies of more than $50\%$ of its mean value indicating how heterogeneous are the dynamics. %
In addition, as the droplets in the $Re=10,000$ case have a relative large initial Stokes number, they tend to stay longer within the nearly saturated turbulent jet core, which can result in an additional delay for the vaporization process in the near field. 
%
\begin{figure}[!hb]
    \centering
    \begin{subfigure}[b]{1.\linewidth}
        \centering
        \includegraphics[scale=0.35]{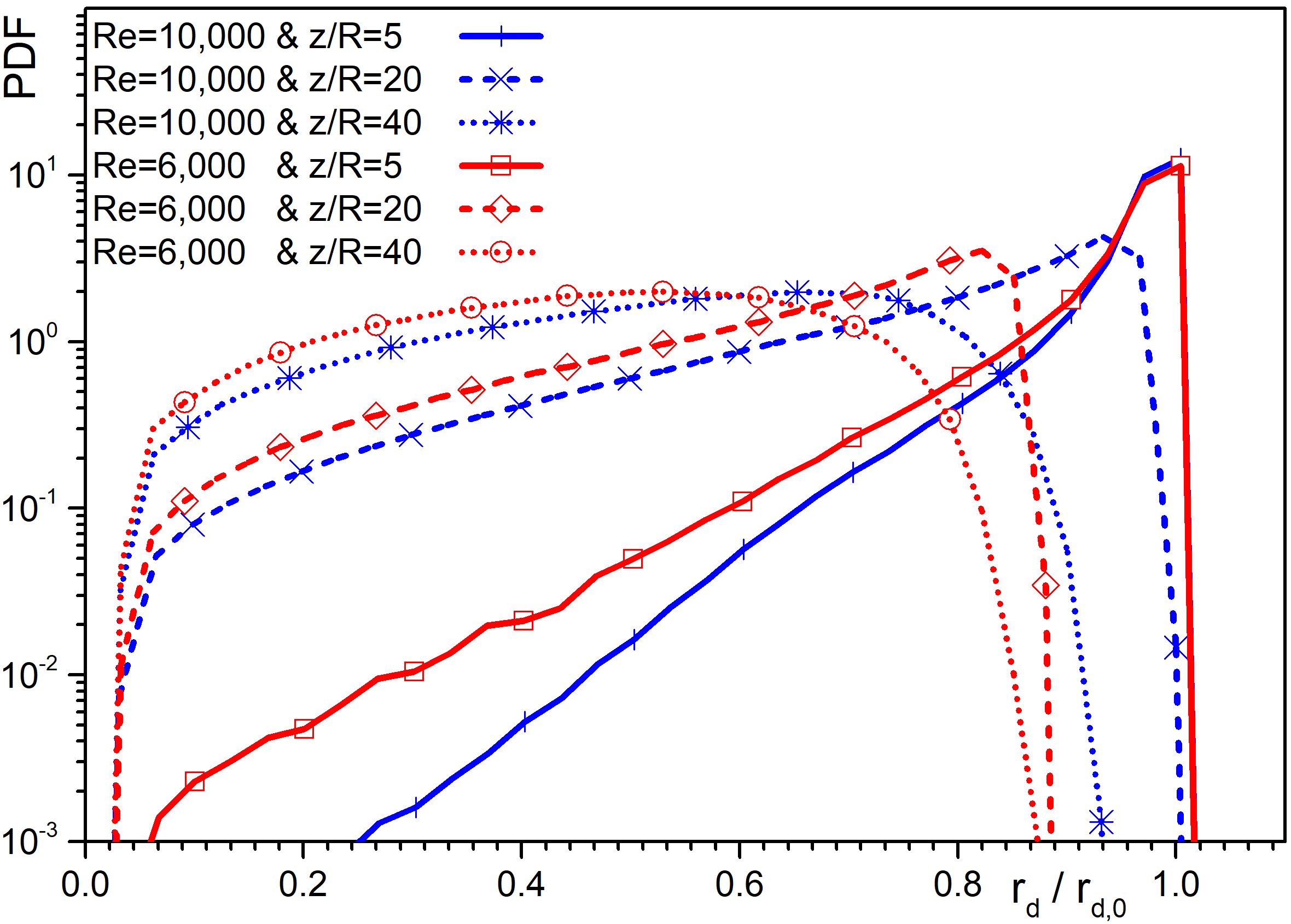}
        \caption{}
        \label{fig:PDF_a}
    \end{subfigure}
    \begin{subfigure}[b]{1.\linewidth}
        \centering
        \includegraphics[scale=0.35]{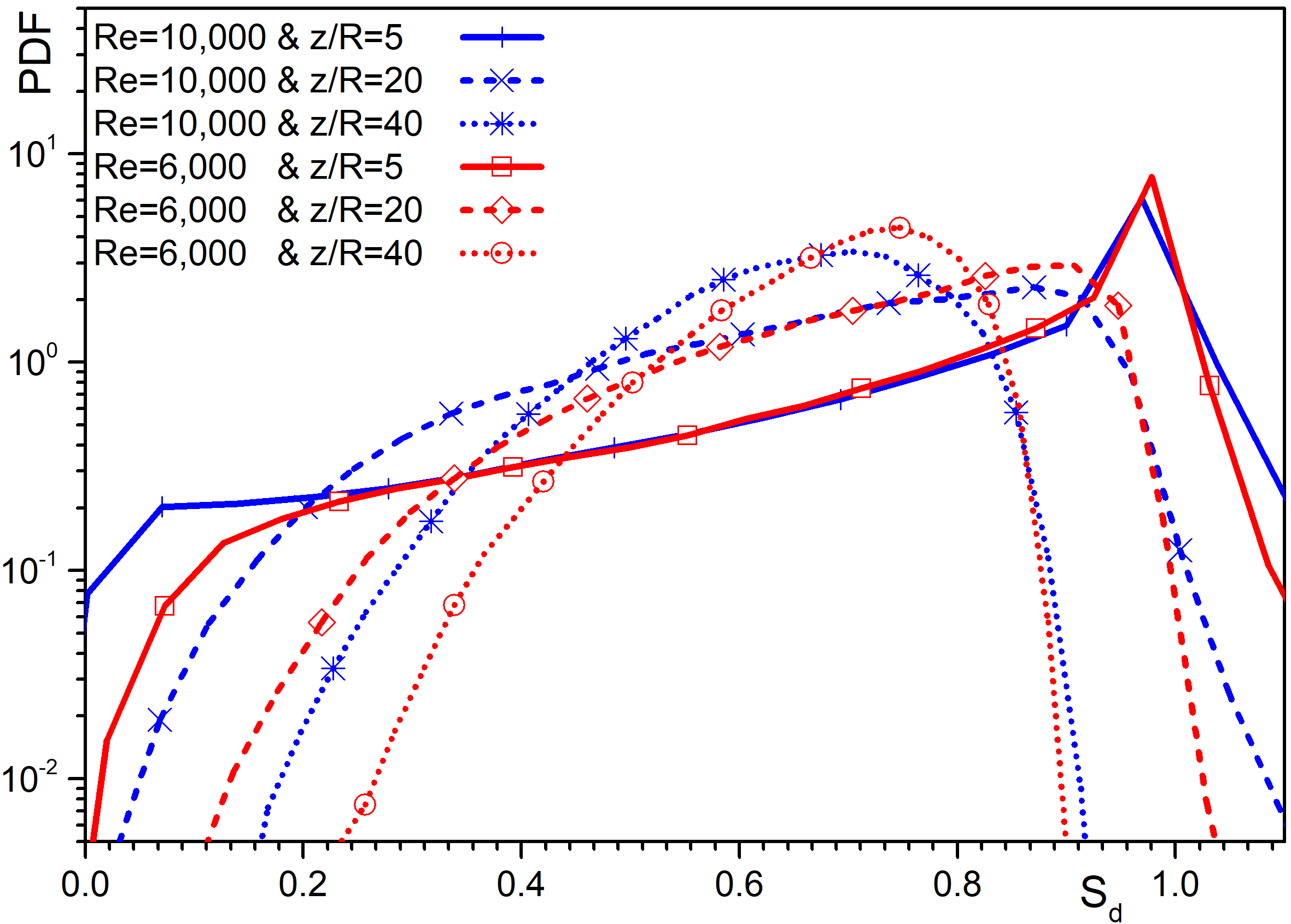}
        \caption{}
        \label{fig:PDF_b}
    \end{subfigure}
    \begin{subfigure}[b]{1.\linewidth}
        \centering
        \includegraphics[scale=0.35]{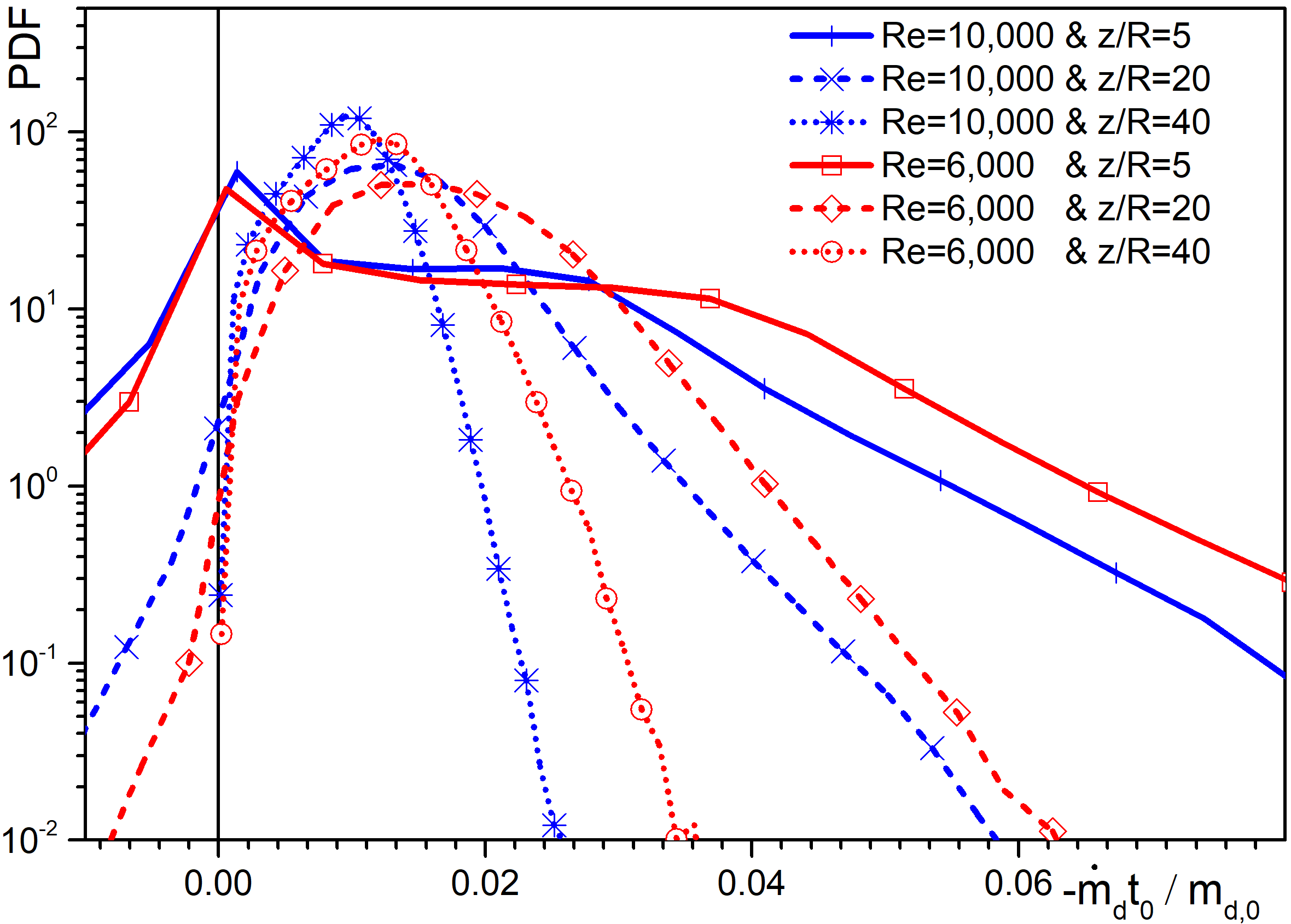}
        \caption{}
        \label{fig:PDF_c}
    \end{subfigure}
\caption{PDF of Lagrangian observables for different Reynolds numbers and $z/R$ positions. (a) PDF of non-dimensional droplet radius, ${r_d}/{r_{d,0}}$, where $r_{d,0}$ is the initial radius of the injected droplets. (b) PDF of the saturation field at droplet surface, $S_d=Y_v / Y_{v,s}$, where $Y_{v,s}=Y_{v,s}(T_d,p)$ is the vapor mass fraction at saturation computed as a function of the droplet actual temperature and the carrier phase thermodynamics pressure,$p_0$. $Y_v$ is the actual vapor mass fraction in the carrier gaseous mixture evaluated at droplet position. (c) PDF of non-dimensional droplet vaporization rate, ${{-\dot{m_d}}t_0}/{r_{d,0}}$, where $t_0$ and $r_{d,0}$ are the the reference time scale of the jet and initial radius of the injected droplets.}
\label{fig:PDF}
\end{figure}
%
Figure~\ref{fig:PDF}(a) compares the probability density function (PDF) of the droplet radius at different $z/R$ positions for the two Reynolds numbers considered. We note that all PDF profiles end sharply. This is due to the fact that droplets with a radius $r_d < 0.5 \mu m$ are removed from the simulation, because of the numerical stability of the evolution of tiny droplets (more than $99.9\%$ of the liquid mass is evaporated before). It is worth remarking how fast the turbulent spray is in promoting polydispersity. %
Though starting from an identical monodisperse status, the droplet radius distribution  spreads over a wide range at $Re=6,000$ even at five jet radii from the inlet, whereas a slightly narrower  PDF is observed at $Re=10,000$. %
Further downstream, i.e.\ $z/R=20$ and $z/R=40$, a similar qualitative behavior can be observed: the  flat PDFs are related to a nearly constant probability to find droplets of arbitrary sizes. %
We attribute this fast widening of the droplet size spectrum to the strong inhomogeneous conditions of local vapour concentration. In the mixing layer, the entrainment process originates spatial regions with very low levels of saturation, which promote a fast droplet vaporization. %
Concurrently, because of clustering, clouds of droplets originating in the jet core maintain nearly zero vaporization rates for long time. %
Focusing on panel (b) of figure~\ref{fig:PDF}, it is possible to appreciate how flat is the PDF of the saturation level of the droplet atmospheres. This reflects in a strongly varying droplet vaporization rate, whose PDF is shown in panel (c). The droplet vaporization rate is slightly higher for the low Reynolds number case, consistently with the slower jet evaporation length and previous discussion on the $Re$ number effect. Finally, we aim to highlight that in the near and intermediate fields some rare condensation events occur. When the droplet evaporation proceeds too fast, which is more common in near-field region close to mixing layer, droplet temperature decreases resulting in the appearing of saturation conditions which can block the evaporation even in a region without high values of vapor concentration. Then, if the droplets move into a higher vapor concentration zone due to  its inertia,  supersaturation conditions could arise. These inertial effects are more worthy of attention for droplets with a higher Stokes number, which is the reason of the relatively broader distribution range of the saturation shown in the $Re = 10,000$ case.
%
\begin{figure}[!b]
    \centering
    \begin{subfigure}[b]{1.\linewidth}
        \centering
        \includegraphics[width=0.98\linewidth]{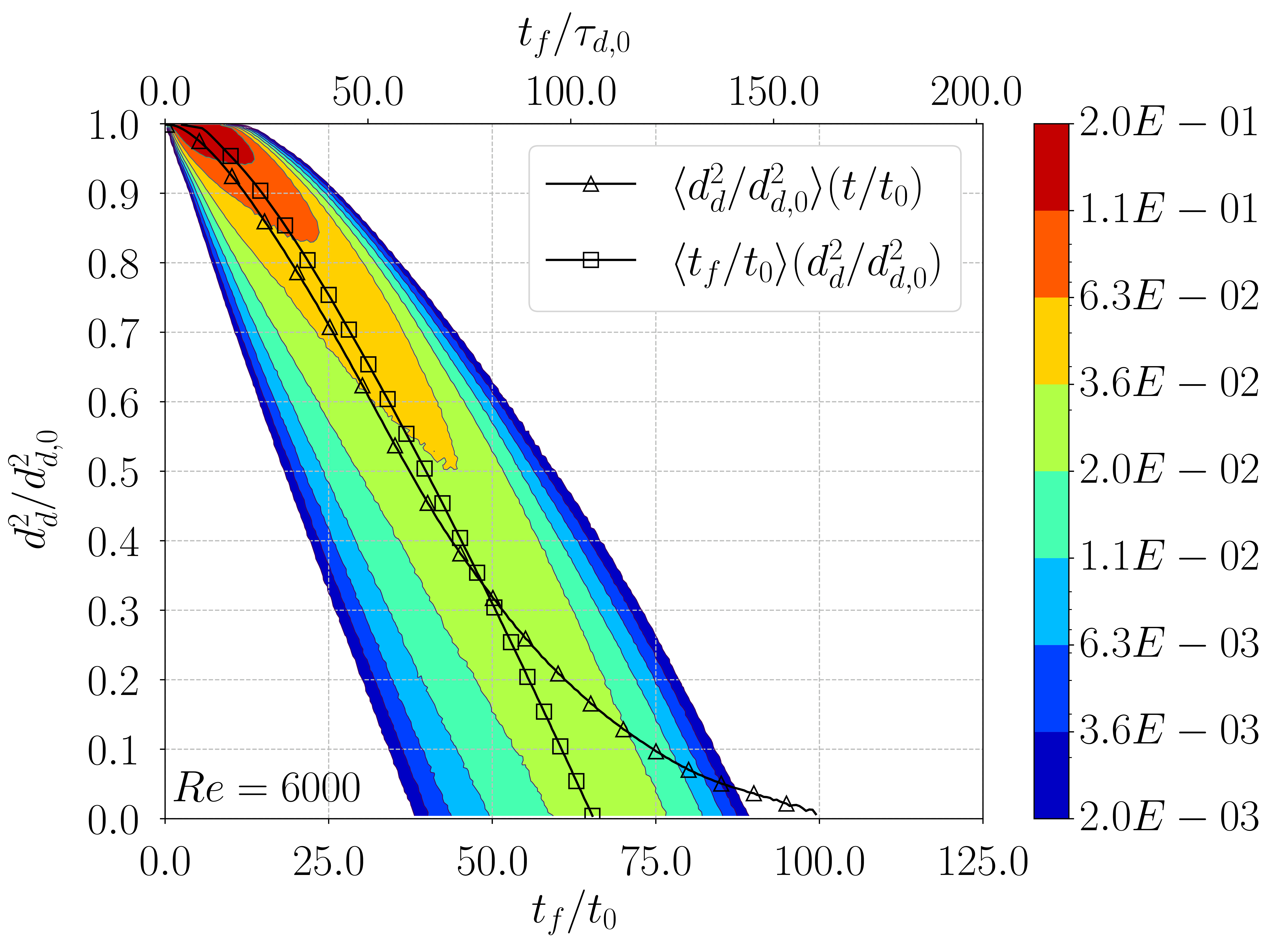}
        \caption{}
        \label{fig:JPDF_d2t_a}
    \end{subfigure}
    \begin{subfigure}[b]{1.\linewidth}
        \centering
        \includegraphics[width=0.98\linewidth]{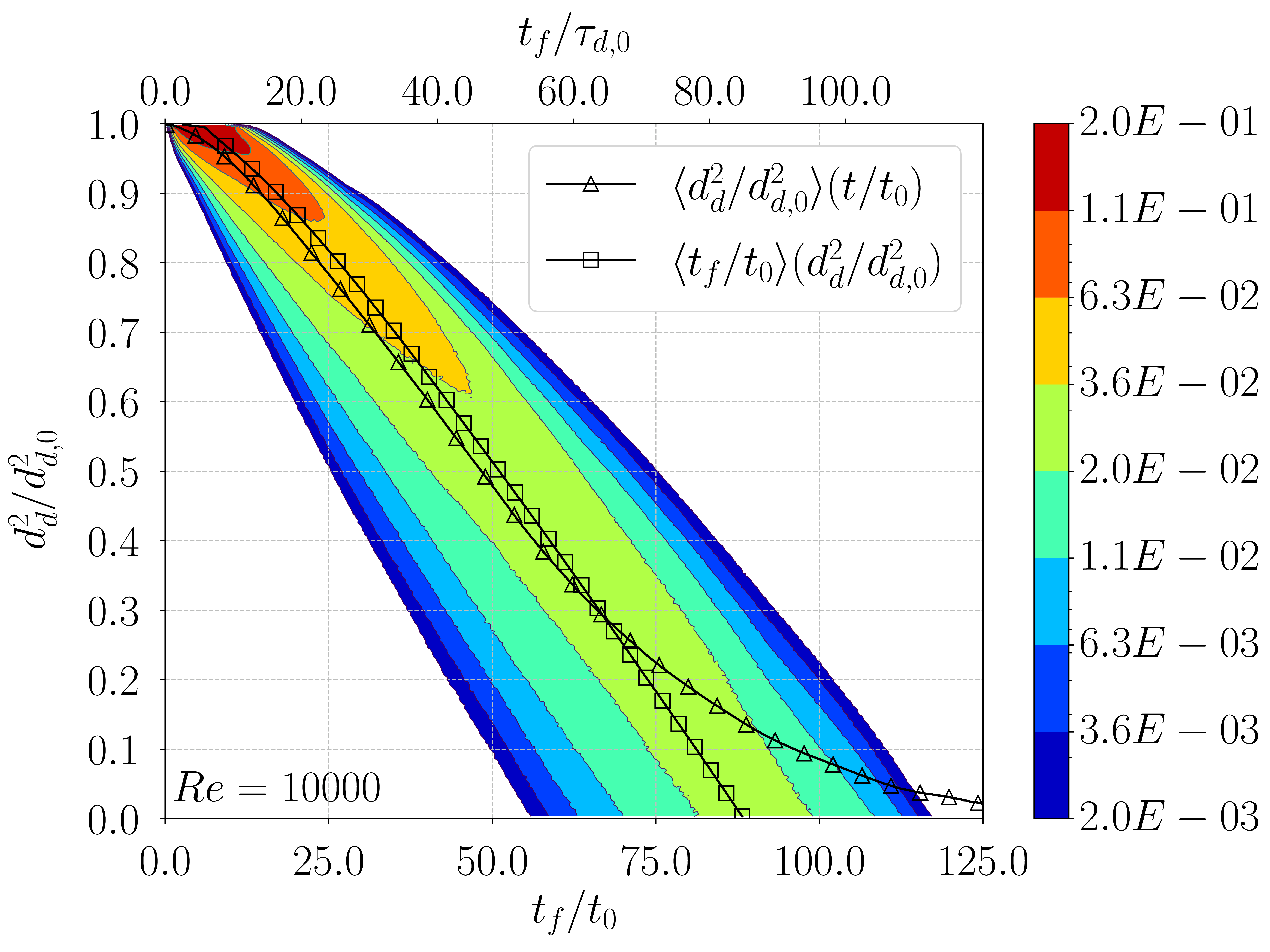}
        \caption{}
        \label{fig:JPDF_d2t_b}
    \end{subfigure}
\caption{Joint probability density function of droplet square diameter and flight time, $D2F(t_f,d^2_d)$, for for different Reynolds numbers. The mean square droplet diameter conditioned to the flight time (triangle symbol), $\langle {d^2_d}/{d^2_{d,0}}\rangle ({t}/{t_0})$, and the mean droplet flight time conditioned to the square droplet diameter (square symbol), $\langle {t_f}/{t_0}\rangle ({d^2_d}/{d^2_{d,0}})$. In the upper abscissa the time is normalized by the initial particle relaxation time, whereas the lower axis provides the non-dimensional time, $t_f/t_0$.}
\label{fig:JPDF_d2t}
\end{figure}
%
The strongly polydisperse behavior shown by the evaporating droplets affects the vaporization length and time of the droplets within the spray. %
\par
To predict the temporal evolution of the droplet dynamics, the {\it d-square} law is often adopted. In this context, a linear behavior for the square diameter of each droplet is expected,
\begin{align}
&\frac{d^2_d}{d^2_{d,0}} \simeq 1 - k \frac{t_f}{t_0},\label{eq:drop_d2law}\\
&k = \frac{Sh}{Sc}\frac{\rho}{\rho_l}\frac{1}{Re}\left(\frac{R}{r_{d,0}}\right)^2ln(1+B_m),
\end{align}
where, in many practical and experimental application, the constant $k$ can only be estimated based on the reference environmental state, e.g.\ bulk vapor concentration. We provide in Fig.~\ref{fig:JPDF_d2t} the joint probability density function of the droplet square diameter and flight time, $D2F(t_f,d^2_d)$. As expected, droplets do not follow a linear time behavior for the square diameter as predicted by the {\it d-square} law, otherwise all data should collapse on a single straight line. In the same figure, we also report the average droplet square diameter conditioned to the flight time, $\langle d^2_d / d^2_{d,0}\rangle(t / t_0)$, and the droplets mean flight time conditioned to the square droplet diameter, $\langle t_f/t_0\rangle (d^2_d/d^2_{d,0})$, for both Reynolds number cases. These two quantities have been directly extracted from the JPDF, $D2F$, as:
\begin{equation} 
    \langle d^2_d\rangle (t_f) = \frac{\int^\infty_0 d^2_d\ \ D2F(t_f,d^2_d)\ d(d^2_d)}{\int^\infty_0\  D2F(t_f,d^2_d)\ d(d^2_d)}
    \label{eq:d2}
\end{equation}
\begin{equation} 
    \langle t_f\rangle(d^2_d) = \frac{\int^\infty_0 t_f\ \ D2F(t_f,d^2_d)\ d(t_f)}{\int^\infty_0 \ D2F(t_f,d^2_d)\ d(t_f)}.
    \label{eq:ft}
\end{equation}
Although the two quantities are strictly related, their significance is different. The mean square diameter, eq.~\eqref{eq:d2}, provides the mean square diameter at a fixed flight time, $i.e.$ after a fixed time from droplet injection. The latter, the mean flight time at fixed mean square diameter, highlights the amount of time needed, in average, for a droplet to reach a fixed diameter. In both the considered cases, the two observables assume similar values for relatively large droplets ($d^2_d / d^2_{d,0}>0.25$), while strongly different for small droplets. The mean flight time conditioned to a droplet diameter shows a finite average time for a full evaporation, i.e.\ $d^2_d / d^2_{d,0}\rightarrow 0$: We will name it mean evaporation flight time, $t^e_f$. Consistently with previous findings, the mean evaporation flight time increases with the  Reynolds number when scaled with the advection reference time, being $t^e_f/t_0\simeq 66$ and $t^e_f/t_0\simeq 89$, for $Re=6,000$ and $Re=10,000$ respectively. We also note that, by scaling the mean flight time for the full droplet evaporation with the droplet initial relaxation time, the two cases show closer values, $t^e_f/\tau_{d,0}\simeq 106$ and $t^e_f/\tau_{d,0}\simeq 86$, with a slightly faster evaporation rate for the higher Reynolds number case. This behavior confirms that, by increasing the Reynolds number, two different competing effects are present: a stronger turbulence intensity which tends to fasten the evaporation rate and a higher droplet relative inertia which tends to slow down the process. We  highlight how different are the behavior of the two curves, Eq.~\eqref{eq:d2} versus Eq.~\eqref{eq:ft}, for small droplet sizes. We note that for time longer than the mean evaporation flight time, $t^e_f$ , we still have a finite average droplet diameter. This peculiar behavior can be explained considering that droplets in clusters require very long time to evaporate, hence some droplets are able to survive for very long time displaying a finite average diameter. This last aspect could be crucial in the dispersion modeling of infectious droplets~\citep{Balachandar.2020}. 
%
\begin{figure}[!b]
    \centering
    \begin{subfigure}[b]{1.\linewidth}
        \centering
        \includegraphics[width=0.98\linewidth]{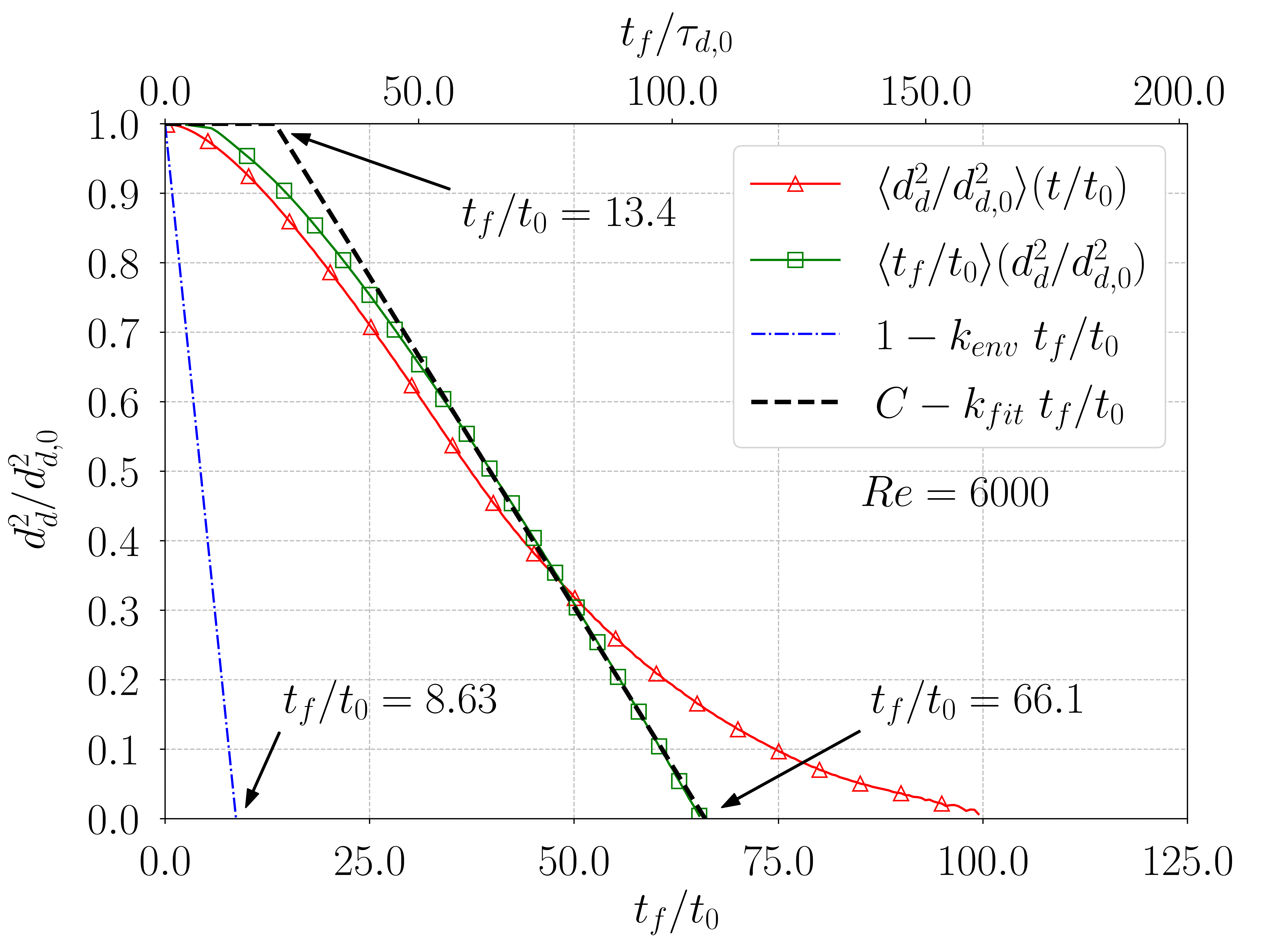}
        \caption{}
        \label{fig:D2law_a}
    \end{subfigure}
    \begin{subfigure}[b]{1.\linewidth}
        \centering
        \includegraphics[width=0.98\linewidth]{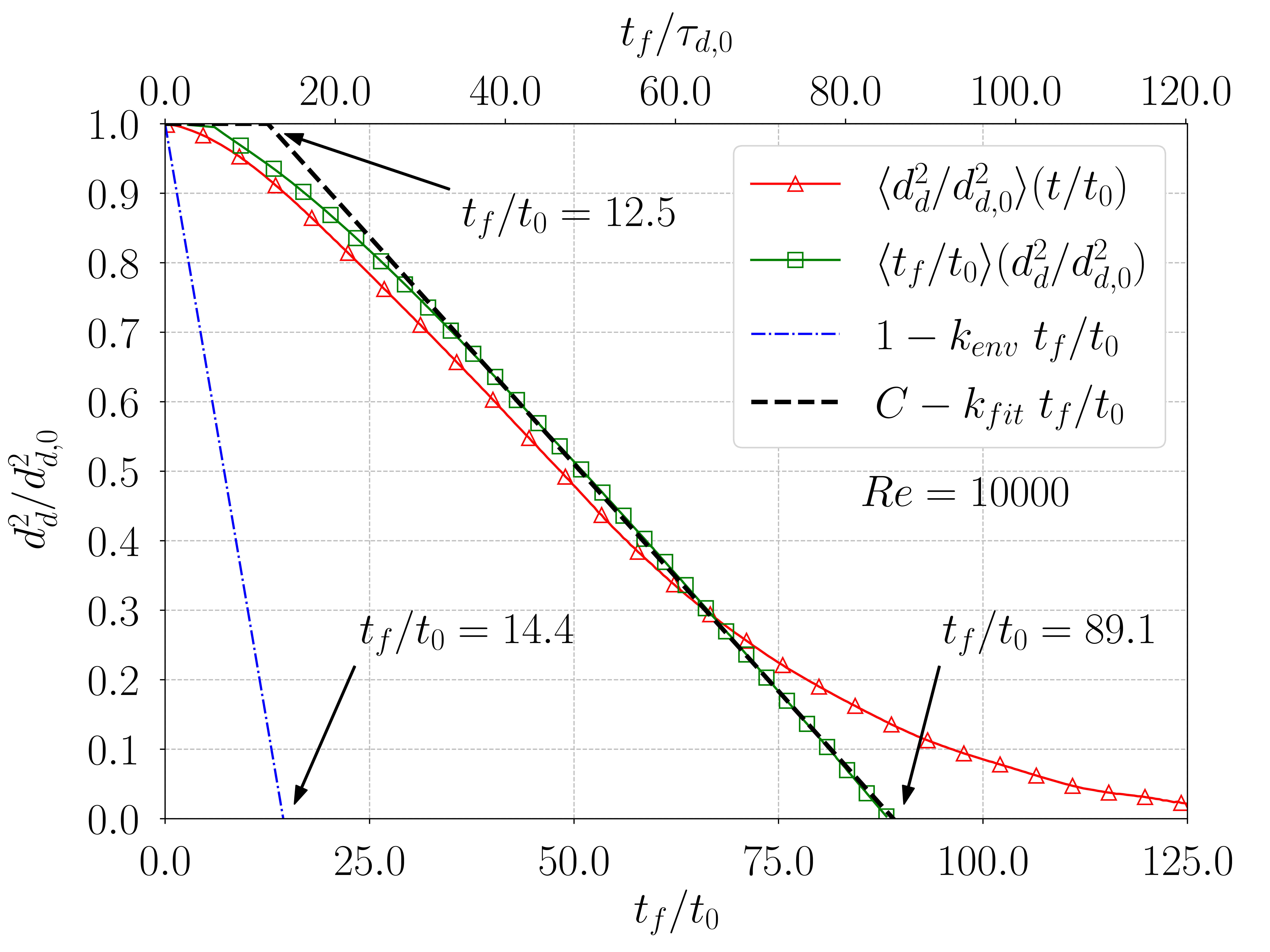}
        \caption{}
        \label{fig:D2law_b}
    \end{subfigure}
\caption{Mean square droplet diameter conditioned to the flight time (triangle symbol), $\langle {d^2_d}/{d^2_{d,0}}\rangle ({t}/{t_0})$, and the mean droplet flight time conditioned to the square droplet diameter (square symbol), $\langle {t_f}/{t_0}\rangle ({d^2_d}/{d^2_{d,0}})$. In the upper axis the time is normalized by the initial particle relaxation time, $\tau_{d,0}$, whereas in the lower one by the advecting time, $t_0$. The blue line shows the evolution of droplet square diameter predicted by employing the {\it d-square} law provided in Eq.~\eqref{eq:drop_d2law} and by estimating the $k$ constant via the bulk environmental conditions. The black, dashed line provides the linear fitting of $\langle {t_f}/{t_0}\rangle ({d^2_d}/{d^2_{d,0}})$ restricted to $t/t_0>30$.}
\label{fig:D2law}
\end{figure}
%
To better characterize how the {\it d-square} law is able to describe present cases we provide in figure~\ref{fig:D2law} the evolution of droplet square diameter predicted by the {\it d-square} law, eq.~\eqref{eq:drop_d2law}, estimating the $k$ constant via the bulk environmental conditions, $k_{env}$, together with $\langle {d^2_d}/{d^2_{d,0}}\rangle ({t}/{t_0})$ and $\langle {t_f}/{t_0}\rangle ({d^2_d}/{d^2_{d,0}})$. %
The estimates for the two cases in dimensionless units are $k_{env}\simeq0.115$ and $k_{env}\simeq0.069$ %
for $Re=6,000$ and $Re=10,000$, respectively. %
In both cases the predicted droplet evaporation time is much lower than the actual one, $i.e.$\ the estimate of the evaporation rate is much faster than the actual one. We attribute this disagreement to the effects of clustering and preferential sampling of vapor concentration previously discussed,  that is not accounted by the {\it d-square} law. Since the mean flight time, eq.~\eqref{eq:ft}, shows a linear behavior for small droplets and long time, we provide also the linear fitting in the form $\langle {t_f}/{t_0}\rangle ({d^2_d}/{d^2_{d,0}})\simeq C-k_{fit}(t_f/t_0)$  restricted to $t_f/t_0>30$. %
In this case the fitting constants are $C\simeq 1.25$ and $k_{fit}\simeq 0.0190$ for $Re=6000$ whereas $C\simeq 1.164$ and $k_{fit}\simeq 0.0131$ for $Re=10000$. %
Using these parameters we can reformulate an effective {\it d-square} law as we assume that droplets do not evaporate for some time after the injection when they travel in the saturated core and then follow an effective {\it d-square} law. The initial time when droplets are assumed not evaporating is similar for the two cases, i.e.\ $\sim 13t_0$, which corresponds to a traveled distance around 13 jet radii. %
Then the two estimated values of $k_{fit}$ appear six times lower than than those estimated at environemental conditions, $k_{env}$, for both cases. %
The large discrepancies between the values of $k_{env}$ and $k_{fit}$ further confirm the crucial importance of improving bulk spray dispersion models accounting for the complex dynamics arising from the strong inhomogeneity of droplet and vapor mass fraction distribution in turbulent jet-sprays.

\section{CONCLUSION}

A direct numerical simulation has been conducted to investigate the motion and evaporation  of inertial droplets in a high-Reynolds-number turbulent jet-spray ($Re=10,000$). %
The Eulerian-Lagrangian point-droplet framework combined with a two way coupling concept, including the mass, momentum and energy exchanges between two phases, is considered for the present acetone-air spray. %
Monodisperse acetone droplets are continuously injected within the turbulent gaseous phase at a bulk Reynolds number $ Re = 2 U_0 R/\nu = 10,000$. Then, a systematic and comprehensive dataset of both the instantaneous and mean fields from Eulerian and Lagrangian observables are collected, analyzed and compared to a  dataset at lower Reynolds number, $Re=6,000$, which from physical point of view corresponds to a lower jet velocity, \citep{DallaBarba.2018}. This comparison allowed to study the $Re$-number effect on the turbulent  evaporation of droplets in dilute sprays.
Both cases show a strong evaporation rate in the mixing layer where the entrainment of dry air dilutes the saturated jet core. However, a longer evaporation length is observed for the higher Reynolds number case. This feature is related to the slower average vaporization rate observed at $Re=10,000$ when scaled by the jet advection reference time scale, $t_0=R/U_0$. This slower rate is attributed to the higher droplet relative inertia (Stokes number) which slows-down the vapour-liquid mass exchange. 

An intense droplet clustering is apparent in both cases. This originates in the mixing layer of the near field and propagates downstream. In the mixing layer, we identify two states separated by the turbulent/non-turbulent interface: the entrained dry air depleted of droplets and saturated gas clouds full of droplets coming from the saturated core. 
Since droplets entering in dry entrained regions fast evaporate, while aggregates in vapor-saturated clouds cannot evaporate, these dynamics tends to intensify the clustering in the downstream evolution. 
In addition, in the far-field evolution small-scale clustering is also promoted by a Kolmogorov Stokes number order one for present cases. 
These dynamics strongly impact the Lagrangian evolution of the droplets creating very different histories. The droplet size spectrum becomes extremely wide even starting from a monodisperse distribution. This widening is associated to also wide PDF of the evaporation rate and saturation level felt by the droplets. These findings could play a role in explaining the fast widening of the droplet spectrum in warm clouds~\cite{sardina2015continuous}.
Finally, we aim to comment on the accuracy of the {\it d-square} law in approximating the droplet evolution for present cases. 
First, we note that the actual droplet evaporation time varies of more than $50\%$ of its mean value indicating that actual droplet history differs from the {\it d-square} law. 
Second, while the mean evaporation (flight) time, $t^e_f$, could be approximated by a {\it d-square} law, the slope $k$ appears much lower than the value extracted from environmental conditions. In addition, we note that after time much longer than $t^e_f$ a full evaporation of droplets does not occur, highlighted by a finite mean droplet size (square diameter).  
This peculiar aspect has been attributed to the clustering dynamics. Aggregate of droplets with a nearly saturated atmosphere may survive for longer time before a full evaporation. Hence, some droplets may show relatively large sizes for long time and distances. This behavior should be considered in the modeling of the dispersion of infectious droplets since some droplets persist much longer than what expected using the {\it d-square} law especially if based on environmental conditions. 
Perspective of this study is to assess the ability of Large-Eddy-Simulation models to capture this complex dynamics at relatively high Reynolds number. 

\section{ACKNOWLEDGEMENT}
This study has been funded by the research project promoted by  China Scholarship Council (CSC)\ Grant\ \#20180625\linebreak0023. We also thank CINECA for providing computational resources via the ISCRA C project: HiReS (HP10C3YNLC).

\bibliographystyle{cas-model2-names}

\bibliography{main}

\end{document}